\documentclass{aa}  

\usepackage{graphics,graphicx}
\usepackage{xcolor}
\usepackage{amsmath}
\usepackage{amssymb}
\usepackage{braket}
\usepackage{wrapfig}
\usepackage{caption}
\usepackage{subcaption}

\begin{document}
\title{Three-dimensional magnetic field imaging of protoplanetary disks using Zeeman broadening and linear polarization observations}
\titlerunning{Zeeman effect in protoplanetary disks}
\author{Boy Lankhaar\inst{1,2} \and Richard Teague\inst{3}}
\institute{\inst{1}Department of Space, Earth and Environment, Chalmers University of Technology, Onsala Space Observatory, 439 92 Onsala, Sweden;
\email{boy.lankhaar@chalmers.se} \\
\inst{2}Leiden Observatory, Leiden University, Post Office Box 9513, 2300 RA Leiden, Netherlands \\
\inst{3}Department of Earth, Atmospheric, and Planetary Sciences, Massachusetts Institute of Technology, Cambridge, MA 02139, USA 
}

\date{}

\abstract
   {Magnetic fields are predicted to have a crucial impact on the structure, evolution and chemistry of protoplanetary disks. However, a direct detection of the magnetic field towards these objects has yet to be achieved.}
   {In order to characterize protoplanetary disk magnetic fields, we investigate the impact of the Zeeman effect on the (polarized) radiative transfer of emission from paramagnetic molecules excited in protoplanetary disks.}
   {While the effects of the Zeeman effect are commonly studied in the circular polarization of spectral lines, we perform a comprehensive modeling also of the Zeeman-induced broadening of spectral lines and their linear polarization. We develop simplified radiative transfer models adapted to protoplanetary disks, which we compare to full three-dimensional polarized radiative transfer simulations.}
   {We find that the radiative transfer of circular polarization is heavily affected by the expected polarity-change of the magnetic field between opposite sides of the disk. In contrast, Zeeman broadening and linear polarization are relatively unaffected by this sign change due to their quadratic dependence on the magnetic field. We can match our simplified radiative transfer models to full polarization modeling with high fidelity, which in turn allows us to prescribe straight-forward methods to extract magnetic field information from Zeeman broadening and linear polarization observations.}
   {We find that Zeeman broadening and linear polarization observations are highly advantageous methods to characterize protoplanetary disk magnetic fields as they are both sensitive probes of the magnetic field and are marginally affected by any sign change of the disk magnetic field. Applying our results to existing circular polarization observations of protoplanetary disk specral lines suggests that the current upper limits on the toroidal magnetic field strengths have to be raised.}
   {}

\keywords{}

\maketitle

\section{Introduction}
The formation of stars is associated with a protoplanetary disk in the late evolutionary stages. The protoplanetary disk is composed of gas and dust and orbits a young star that is actively accreting mass through it. Planets are believed to be formed in the protoplanetary disk; a process that crucially depends on the disk structure, chemistry and evolution \citep{armitage:11}. 

The disk structure and evolution sensitively depends on the dominant accretion mechanism. Accretion requires the transport of angular momentum through the disk, which is facilitated by turbulence-enhanced viscosity, or through magnetic fields, that may either transport angular momentum radially through magnetic stresses or vertically through the launching of a disk wind \citep{balbus:91, nelson:13, lesur:10, marcus:15, wardle:93, shakura:76, armitage:19, bai:13}. The lack of observational signatures of turbulence in protoplanetary disks catalyzed substantial progress in numerical modeling, which put forth disk winds as the dominant accretion mechanism \citep{flaherty:15, flaherty:18, teague:16, bai:16, lesur:21}. The details of the disk wind crucially depend on the strength and morphology of the disk magnetic field \citep[see, e.g.][]{bai:16}. Additionally, magnetic fields advected through the disk wind affect disk chemistry through the shielding of cosmic rays \citep{cleeves:13}. 

Despite its importance, a direct detection of the large scale magnetic field that permeates a protoplanetary disk has not yet been achieved despite several searches \citep{vlemmings:19, harrison:21}. Magnetic fields in molecular gas are commonly observed through the detection of linear polarization in dust continuum emission \citep{andersson:15, pattle:22} or spectral lines, either in thermal emission \citep{goldreich:81, lankhaar:20b}, or in maser emission \citep{crutcher:19}, or through the detection of the Zeeman effect in the circularly polarized spectral line emission of paramagnetic molecules \citep{crutcher:19}. Applied to the protoplanetary disk, these methods of magnetic field detection have had limited success. While dust continuum polarization is readily detected with ALMA, radiation scattering affects the polarization signature, making interpretation of these observations for their magnetic field information difficult \citep{kataoka:15, stephens:17}. The linear polarization of spectral lines has been tentatively detected in $^{12}$CO and $^{13}$CO emission, but at low degrees of polarization due to the high density of these regions \citep{stephens:20, teague:21, lankhaar:20a, lankhaar:22a}. Circular polarization observations have been able to put stringent constraints on the magnetic field strengths of TW Hya \citep{vlemmings:19} and AS 209 \citep{harrison:21}, but failed to detect magnetic fields directly due to the high sensitivity requirements of circular polarization observations. 
%

Zeeman observations in protoplanetary disks through circular polarization have high sensitivity requirements for two reasons. First, observation of circularly polarized emission of spectral lines is inherently affected by instrumental effects that require strong signals, as well as excellent atmospheric conditions, for proper calibration. Second, circular polarization observations are sensitive to the line-of-sight component of the magnetic field. This is problematic, as protoplanetary disks are believed to be permeated by a magnetic field that is dominated by its toroidal component; $\gtrsim 10$ times stronger than the vertical and radial components. Additionally, the projection of the toroidal and radial components of the magnetic field are expected to change sign between the front and back side of the disk (for more discussion, see text after Eq.~\ref{eq:B_field}). Therefore, for weakly inclined disks it is mainly the weak vertical component of the magnetic field that gives rise to circular polarization. On the other hand, (moderately) inclined disks are affected by complicated radiative transfer effects that cause lower degrees of polarization due to line broadening and the cancellation of the circular polarization due to the sign change of the toroidal magnetic field component \citep{mazzei:20}. 

In this paper, we propose to use the broadening and linear polarization of paramagnetic spectral lines due to the Zeeman effect as tracers of the protoplanetary disk magnetic field. We argue that because Zeeman broadening and linear polarization are dependent on the square of the magnetic field strength \citep{crutcher:93}, they are less sensitive to the sign change of the magnetic field that characterizes protoplanetary disks. Moreover, while circular polarization observations are sensitive to the line-of-sight component of the magnetic field, Zeeman induced linear polarization is sensitive to the plane-of-the-sky component of the magnetic field and Zeeman broadening is dependent on the total magnetic field strength. Therefore, Zeeman broadening and Zeeman induced linear polarization have a high magnetic field sensitivity also for weakly inclined disks. Finally, we argue that from the simultaneous observation of Zeeman induced linear polarization with either Zeeman broadening or the circular polarization, the three-dimensional (3D) direction and strength of the magnetic field may be derived.

This paper is structured as follows. In Section 2, we review the Zeeman effect and its manifestation in spectral lines. We discuss simple radiative transfer models that will assist in the interpretation of intricate 3D radiative transfer modeling. Importantly, we show how to generalize the expressions for Zeeman broadening and linear polarization, derived in the seminal work of \citet{crutcher:93}, to arbitrary transitions. In Section 3, we present 3D simulations of the polarized radiative transfer of CN emission lines that are excited in a protoplanetary disk which is permeated by a magnetic field. We discuss the emergence of circular and linear polarization, as well as the Zeeman broadening, and show how to extract magnetic field information from synthetic observations. In Section 4, we discuss our results before concluding in Section 5.


\section{Theory}
\label{sec:theory}

In this work, we focus on the detection of magnetic fields in protoplanetary disks through the Zeeman effect. In the following, we discuss the Zeeman splitting of spectral lines and their manifestation in the circular polarization, linear polarization, and in the broadening of spectral lines. In particular, linear polarization and line broadening are often overlooked features of Zeeman split spectral lines excited in interstellar matter, but are established tools of stellar magnetic field detection \citep{robinson:80, gray:84, rosen:15, kochukhov:19}. Here, we derive simple relations for the polarization and broadening of spectral lines as a function of the magnetic field strength and orientation. After the review of Zeeman effects, we consider the manifestation of Zeeman effects in the spectral line emission from protoplanetary disks. We describe the radiative transfer in protoplanetary disks in terms of simplified radiative transfer models and discuss the impact of disk inclination and an optically thick dust layer. We conclude this section with a discussion of stacking procedures that may be applied to extract magnetic field information from (polarized) spectral line emission emerging from protoplanetary disks.

\subsection{Zeeman splitting of spectral lines}
\label{sec:theory_zeeman}

Spectral lines split into a manifold of transitions between their magnetic sublevels due to the Zeeman effect. If we consider a transition between two states, with angular momentum $F_1$ for the upper state and $F_2$ for the lower state, transitions between magnetic sublevels are shifted in frequency by,
\begin{align}
\label{eq:zeeman_shift}
\Delta \nu_Z (F_1,m_1,F_2,m_2) = \frac{\mu_B}{h} B \left(g_1 m_1 - g_2 m_2 \right),
\end{align}
which is linearly dependent on the magnetic field $B$. In Eq.~(\ref{eq:zeeman_shift}) $\mu_B$ is the Bohr magneton, $h$ is Planck's constant, $g_1$ and $g_2$ are the $g$-factors of the two states and $m_1$ and $m_2$ are the magnetic quantum numbers that can assume values of $-F_1 (-F_2)$ to $F_1 (F_2)$ in increments of $1$. The transition $\ket{F_1 m_1} \to \ket{F_2 m_2}$ is associated with a relative line-strength $S_{m_1-m_2}(F_1,F_2,m_1)$ \citep[see equation 3.16 in][for a definition of the relative line-strength]{landi:06}. Due to selection rules, only $\Delta m=0$ ($\pi^0$-transitions) and $\Delta m=\pm 1$ ($\sigma^{\pm}$-transitions) transitions are allowed. The groups of $\pi^0$- and $\sigma^{\pm}$-transitions have different opacities for the different polarization modes of the radiation field which are a function of the projection angle, $\theta$, between the line-of-sight direction and the magnetic field direction. For a magnetic field along the line-of-sight ($\cos \theta =1$), $\sigma^{\pm}$-transitions give rise to right- and left-handedly circularly polarized radiation, while for a magnetic field perpendicular to the line-of-sight ($\cos \theta=0$), $\sigma^{\pm}$-transitions give rise to linear polarization perpendicular to the magnetic field direction in the plane-of-the-sky. The $\pi^0$-transitions are suppressed when the magnetic field is oriented along the line-of-sight, while for a magnetic field perpendicular to the line-of-sight, $\pi^0$-transitions give rise to linear polarization parallel to the magnetic field direction in the plane-of-the-sky.

When there is no magnetic field present, the individual transitions between the magnetic sublevels are degenerate and the different propagation properties of the $\sigma^{\pm}-$ and $\pi^0$-transitions for different polarization modes do not come to expression. Through interaction with the magnetic field, the degeneracy is lifted and the $\sigma^{\pm}$- and $\pi^0$-transitions are shifted with respect to each other, resulting in the partial polarization (circular and linear), as well as a broadening of the associated spectral line \citep{landi:06}. Most commonly, the signature of the Zeeman effect is sought in the circular polarization, but in this paper we focus also on the Zeeman signature in the linear polarization and line broadening.

To relate our discussion to the published literature on Zeeman effects, we first discuss the signature of the Zeeman effect in the spectral line circular polarization. The $\sigma^{\pm}$-transitions are oppositely circularly polarized, and therefore a net polarization is produced when these transition-groups are shifted with respect to one another. The groups of $\sigma^{\pm}$-transitions are shifted from the line-center (on average, weighted by intensity), in frequency by,
\begin{align}
\label{eq:zeeman_fo}
\pm \nu_Z &= \sum_{m_1} \Delta \nu_Z(F_1,m_1,F_2,m_1\pm 1) S_{\pm 1}(F_1,F_2,m_1)\nonumber \\ &= \pm zB/2,
\end{align}
where $z$ is the line-specific Zeeman coefficient and $B$ is the magnetic field strength \citep{landi:06, larsson:19, vlemmings:19}. Since the expression of the Zeeman effect in the circular polarization is through a shift of the $\sigma^{\pm}$ line profiles, its relative strength is inversely proportional to the line width. It is therefore helpful to define the Zeeman shift in Doppler units,
\begin{align}
x_Z &= \frac{c \nu_Z}{\nu_0 b} = 0.15 \left( \frac{z}{2\ \mathrm{kHz/mG}}\right)\left(\frac{B}{10\ \mathrm{mG}}\right) \nonumber \\ 
&\times \left(\frac{b}{200\ \mathrm{m/s}}\right)^{-1} \left(\frac{\nu_0}{100 \ \mathrm{GHz}}\right)^{-1},
\end{align}
where $\nu_0$ is the line frequency and $b$ is the Doppler b parameter which describes the line broadening and usually has a contribution both from the thermal broadening, $b_{\mathrm{thermal}} = \sqrt{2kT/m}$, where $m$ is the particle mass and $T$ the temperature, and the turbulent broadening $b_{\mathrm{turb}}$. In addition to a shift between the transition groups, there is also a spread in frequency within the $\pi^0$- and the $\sigma^{\pm}$-transition groups that causes line broadening and impacts the production of linear polarization. For Zeeman shifts that are weaker than the Doppler broadening, the effects of intragroup broadening may be incorporated in the polarized radiative transfer perturbatively \citep[for a detailed discussion, see Chapter 9 of][and Appendix \ref{sec:ap_A}]{landi:06}. Taking this into account, we derive in Appendix \ref{sec:ap_A} the polarized radiative transfer equation, which we report here only in the optically thin limit,
\begin{subequations}
\label{eq:zeeman_thin}
\begin{align}
I_{\nu} &= S_{\nu_0} \tau_{\nu_0} \left[\bar{\phi} (x) +x_Z^2\left(\frac{\bar{Q}}{4} - \frac{\Delta Q}{4}\cos^2 \theta \right) \bar{\phi}''(x)  \right],\\
Q_{\nu} &= S_{\nu_0} \tau_{\nu_0} x_Z^2 \Delta Q \frac{\sin^2 \theta \cos 2\eta}{4} \bar{\phi}''(x), \\
U_{\nu} &= S_{\nu_0} \tau_{\nu_0} x_Z^2 \Delta Q \frac{\sin^2 \theta \sin 2\eta}{4} \bar{\phi}''(x), \\
V_{\nu} &= S_{\nu_0} \tau_{\nu_0} x_Z \cos \theta \bar{\phi}'(x),
\end{align}
\end{subequations}
where $S_{\nu_0}$ is the source function of a spectral line at frequency $\nu_0$, $\tau_{\nu_0}$ is its line optical depth at the line center, $\eta$ is the angle between the magnetic field projected on the plane of the sky and the plane of the sky Northern direction, and $\bar{\phi}'(x)$ and $\bar{\phi}''(x)$ indicate the first- and second-derivative of the normalized line profile, $\bar{\phi}(x)=e^{-x^2}/\sqrt{\pi}$, with respect to the Doppler unit, $x=\frac{c(\nu-\nu_0)/\nu_0}{b}$. We refer to the angle $\eta$ as the magnetic field position angle in the remainder of this paper. The dimensionless coefficients $\bar{Q}$ and $\Delta Q$ account for the contribution of the intragroup broadening to the total Zeeman broadening and the emergence of linear polarization. They are defined, $\bar{Q}=Q^{0} + Q^{\pm}$ and $\Delta Q = Q^0-Q^{\pm}$, where,
\begin{subequations}
\label{eq:zeeman_so}
\begin{align}
\nu_Z^2 Q^0 &= \sum_{m_1} \left[\Delta \nu_Z(F_1,m_1,F_2,m_1)\right]^2 S_{0}(F_1,F_2,m_1),  \\
\nu_Z^2 Q^\pm &= \sum_{m_1} \left[\Delta \nu_Z(F_1,m_1,F_2,m_1\pm 1)\right]^2 S_{\pm 1}(F_1,F_2,m_1), 
\end{align}
\end{subequations}
are readily computed from the level-specific $g$-factors within a transition. We have tabulated them for CN (sub-)millimeter transitions in Table~\ref{tab:CN_mol}. 

\begin{figure}[h]
\centering
\includegraphics[width=0.49\textwidth]{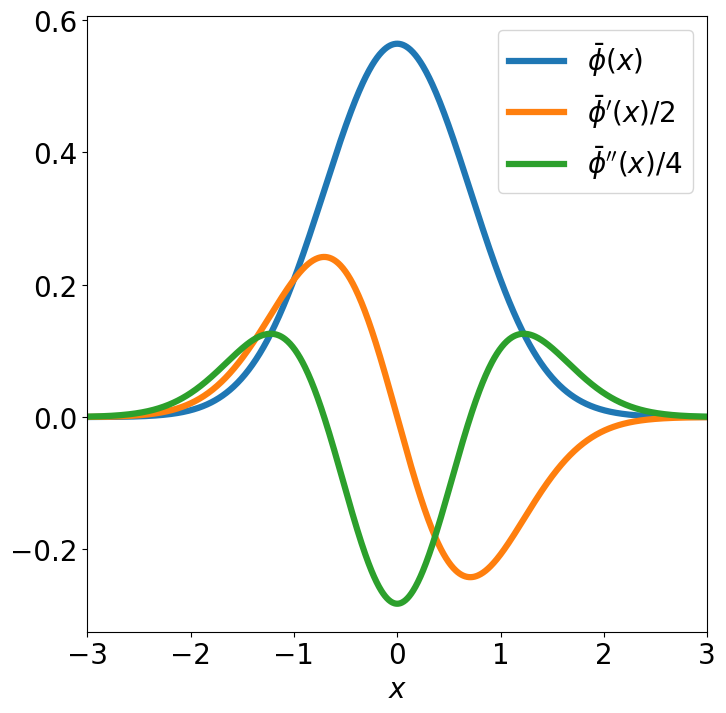}
\caption{Normalized Gaussian line profile and its first two derivatives plotted against the Doppler unit.}
\label{fig:line_profiles}
\end{figure}

In Fig.~(\ref{fig:line_profiles}), we plot the normalized line profile and its derivatives. Comparing the Zeeman term in the total intensity in Eq.~(\ref{eq:zeeman_thin}) to the second-derivative line profile, it is readily recognized that the Zeeman effect causes a broadening of the total intensity profile. While the term, $\bar{Q}-\Delta Q \cos^2 \theta $, is always positive, $\phi''(x)$ is negative towards the line center and positive in its wings, thus the Zeeman term effectively broadens the line profile. The $\Delta Q$ factor can be positive or negative, depending on the transition, and must be known to relate the linear polarization direction to the magnetic field direction. Also note that as the linear polarization adheres to a spectrum that changes sign towards the line wings (at $x=\pm 2^{-\frac{1}{2}}\approx \pm 0.7$), a spectral resolution that allows for the line profile to be resolved, in addition to a strong Zeeman effect, is required to observe the linear polarization.

\begin{table*}[]
    \caption{Zeeman parameters, frequencies and Einstein coefficients of CN transitions that are relevant to ALMA polarization measurements.}
\label{tab:CN_mol}
    \centering
    \begin{tabular}{l l l l l l c c c c c}
    \hline \hline
    $N$ & $J$ & $F$ & $N'$ & $J'$ & $F'$ & $\nu \ \mathrm{(GHz)}$ & $z \ \mathrm{(Hz/\mu G)}$ & $\bar{Q}$ & $\Delta Q$ & $A \times 10^6 \ \mathrm{(s^{-1})}$ \\
    \hline
$1$ & $0.5$ & $0.5$ & $0$ & $0.5$ & $1.5$ & $113.14416$ & $2.18$ & $1.14$ & $-0.98$ & $10.50$ \\ 
$1$ & $0.5$ & $0.5$ & $0$ & $0.5$ & $0.5$ & $113.12337$ & $-0.62$ & $5.01$ & $3.01$ & $ 1.29$ \\ 
$1$ & $0.5$ & $1.5$ & $0$ & $0.5$ & $1.5$ & $113.19128$ & $0.63$ & $42.84$ & $21.89$ & $ 6.68$ \\ 
$1$ & $0.5$ & $1.5$ & $0$ & $0.5$ & $0.5$ & $113.17049$ & $-0.30$ & $8.42$ & $0.06$ & $ 5.14$ \\ 
$1$ & $1.5$ & $2.5$ & $0$ & $0.5$ & $1.5$ & $113.49097$ & $0.56$ & $4.27$ & $-0.53$ & $11.90$ \\ 
$1$ & $1.5$ & $1.5$ & $0$ & $0.5$ & $1.5$ & $113.50891$ & $1.62$ & $1.26$ & $-0.86$ & $ 5.19$ \\ 
$1$ & $1.5$ & $1.5$ & $0$ & $0.5$ & $0.5$ & $113.48812$ & $2.17$ & $1.97$ & $-0.86$ & $ 6.74$ \\ 
$1$ & $1.5$ & $0.5$ & $0$ & $0.5$ & $1.5$ & $113.52043$ & $1.56$ & $1.28$ & $-0.96$ & $ 1.30$ \\ 
$1$ & $1.5$ & $0.5$ & $0$ & $0.5$ & $0.5$ & $113.49964$ & $0.62$ & $17.07$ & $15.07$ & $10.60$ \\ 
$2$ & $1.5$ & $0.5$ & $1$ & $0.5$ & $0.5$ & $226.66369$ & $-0.62$ & $5.02$ & $3.02$ & $84.60$ \\ 
$2$ & $1.5$ & $0.5$ & $1$ & $0.5$ & $1.5$ & $226.61657$ & $-0.30$ & $8.39$ & $0.06$ & $10.70$ \\ 
$2$ & $1.5$ & $0.5$ & $1$ & $1.5$ & $1.5$ & $226.29894$ & $2.17$ & $1.97$ & $-0.86$ & $ 8.23$ \\ 
$2$ & $1.5$ & $0.5$ & $1$ & $1.5$ & $0.5$ & $226.28742$ & $0.62$ & $17.03$ & $15.03$ & $10.30$ \\ 
$2$ & $1.5$ & $1.5$ & $1$ & $0.5$ & $0.5$ & $226.67931$ & $-1.18$ & $1.65$ & $-0.91$ & $52.70$ \\ 
$2$ & $1.5$ & $1.5$ & $1$ & $0.5$ & $1.5$ & $226.63219$ & $-0.72$ & $1.21$ & $-0.88$ & $42.60$ \\ 
$2$ & $1.5$ & $1.5$ & $1$ & $1.5$ & $2.5$ & $226.33250$ & $2.58$ & $2.04$ & $-0.85$ & $ 4.55$ \\ 
$2$ & $1.5$ & $1.5$ & $1$ & $1.5$ & $1.5$ & $226.31454$ & $0.27$ & $172.48$ & $92.83$ & $ 9.90$ \\ 
$2$ & $1.5$ & $1.5$ & $1$ & $1.5$ & $0.5$ & $226.30304$ & $-1.80$ & $3.08$ & $-0.70$ & $ 4.17$ \\ 
$2$ & $1.5$ & $2.5$ & $1$ & $0.5$ & $1.5$ & $226.65956$ & $-0.71$ & $1.01$ & $-1.00$ & $94.70$ \\ 
$2$ & $1.5$ & $2.5$ & $1$ & $1.5$ & $2.5$ & $226.35987$ & $0.22$ & $423.15$ & $218.01$ & $16.10$ \\ 
$2$ & $1.5$ & $2.5$ & $1$ & $1.5$ & $1.5$ & $226.34193$ & $-2.20$ & $2.58$ & $-0.77$ & $ 3.16$ \\ 
$2$ & $2.5$ & $3.5$ & $1$ & $1.5$ & $2.5$ & $226.87478$ & $0.40$ & $3.52$ & $-0.64$ & $114.00$ \\ 
$2$ & $2.5$ & $2.5$ & $1$ & $0.5$ & $1.5$ & $227.19182$ & $2.20$ & $1.98$ & $-0.86$ & $ 0.00$ \\ 
$2$ & $2.5$ & $2.5$ & $1$ & $1.5$ & $2.5$ & $226.89213$ & $1.06$ & $1.10$ & $-0.95$ & $18.10$ \\ 
$2$ & $2.5$ & $2.5$ & $1$ & $1.5$ & $1.5$ & $226.87419$ & $0.71$ & $1.50$ & $-0.93$ & $96.20$ \\ 
$2$ & $2.5$ & $1.5$ & $1$ & $1.5$ & $2.5$ & $226.90536$ & $0.79$ & $1.60$ & $-0.91$ & $ 1.13$ \\ 
$2$ & $2.5$ & $1.5$ & $1$ & $1.5$ & $1.5$ & $226.88742$ & $1.47$ & $1.05$ & $-0.97$ & $27.30$ \\ 
$2$ & $2.5$ & $1.5$ & $1$ & $1.5$ & $0.5$ & $226.87590$ & $1.18$ & $1.75$ & $-0.89$ & $85.90$ \\ 
$3$ & $2.5$ & $1.5$ & $2$ & $1.5$ & $0.5$ & $340.03541$ & $-0.93$ & $1.28$ & $-0.96$ & $289.00$ \\ 
$3$ & $2.5$ & $1.5$ & $2$ & $1.5$ & $1.5$ & $340.01963$ & $-0.97$ & $1.25$ & $-0.86$ & $92.70$ \\ 
$3$ & $2.5$ & $1.5$ & $2$ & $1.5$ & $2.5$ & $339.99226$ & $-0.33$ & $4.34$ & $-0.52$ & $ 3.89$ \\ 
$3$ & $2.5$ & $1.5$ & $2$ & $2.5$ & $2.5$ & $339.46000$ & $2.57$ & $2.24$ & $-0.82$ & $ 4.33$ \\ 
$3$ & $2.5$ & $1.5$ & $2$ & $2.5$ & $1.5$ & $339.44678$ & $0.22$ & $383.95$ & $208.54$ & $22.60$ \\ 
$3$ & $2.5$ & $2.5$ & $2$ & $1.5$ & $1.5$ & $340.03541$ & $-0.62$ & $1.06$ & $-0.99$ & $323.00$ \\ 
$3$ & $2.5$ & $2.5$ & $2$ & $1.5$ & $2.5$ & $340.00813$ & $-0.69$ & $1.02$ & $-0.99$ & $62.00$ \\ 
$3$ & $2.5$ & $2.5$ & $2$ & $2.5$ & $3.5$ & $339.49321$ & $2.69$ & $2.24$ & $-0.82$ & $ 2.99$ \\ 
$3$ & $2.5$ & $2.5$ & $2$ & $2.5$ & $2.5$ & $339.47590$ & $0.14$ & $964.77$ & $499.00$ & $21.20$ \\ 
$3$ & $2.5$ & $2.5$ & $2$ & $2.5$ & $1.5$ & $339.46264$ & $-2.42$ & $2.63$ & $-0.77$ & $ 2.95$ \\ 
$3$ & $2.5$ & $3.5$ & $2$ & $1.5$ & $2.5$ & $340.03155$ & $-0.45$ & $1.20$ & $-0.97$ & $384.00$ \\ 
$3$ & $2.5$ & $3.5$ & $2$ & $2.5$ & $3.5$ & $339.51664$ & $0.11$ & $1757.21$ & $895.02$ & $25.40$ \\ 
$3$ & $2.5$ & $3.5$ & $2$ & $2.5$ & $2.5$ & $339.49929$ & $-2.52$ & $2.51$ & $-0.78$ & $ 2.33$ \\ 
$3$ & $3.5$ & $4.5$ & $2$ & $2.5$ & $3.5$ & $340.24777$ & $0.31$ & $3.20$ & $-0.69$ & $413.00$ \\ 
$3$ & $3.5$ & $3.5$ & $2$ & $2.5$ & $3.5$ & $340.26495$ & $0.77$ & $1.05$ & $-0.97$ & $33.50$ \\ 
$3$ & $3.5$ & $3.5$ & $2$ & $2.5$ & $2.5$ & $340.24777$ & $0.45$ & $2.15$ & $-0.84$ & $380.00$ \\ 
$3$ & $3.5$ & $2.5$ & $2$ & $2.5$ & $3.5$ & $340.27912$ & $0.51$ & $1.77$ & $-0.89$ & $ 0.93$ \\ 
$3$ & $3.5$ & $2.5$ & $2$ & $2.5$ & $2.5$ & $340.26177$ & $1.01$ & $1.01$ & $-1.00$ & $44.80$ \\ 
$3$ & $3.5$ & $2.5$ & $2$ & $2.5$ & $1.5$ & $340.24854$ & $0.62$ & $2.37$ & $-0.80$ & $367.00$ \\ 
\hline
    \end{tabular}
\end{table*}

We now proceed to discuss the radiative transfer solutions of Eqs.~(\ref{eq:zeeman_thin}) and quantify the impact of the Zeeman effect on the circular polarization, line broadening and linear polarization. The Zeeman effect is most commonly sought in the Stokes $V$ spectrum, i.e.~the circular polarization. To quantify the circular polarization produced through the Zeeman effect, we use the circular polarization fraction,
\begin{align}
p_V = \frac{1}{2}\frac{V_{\mathrm{max}} - V_{\mathrm{min}}}{I_{\mathrm{max}}},
\label{eq:circ_pol_parameter}
\end{align}
where the `$\mathrm{max}$' and `$\mathrm{min}$' subscripts indicate the maximum and minimum value of $V_{\nu}$, respectively. A single optically thin propagation, over a constant magnetic field, yields a polarization fraction
\begin{align}
\label{eq:pV}
p_V^{\mathrm{thin}} \simeq \sqrt{\frac{2}{e}} \ x_Z \cos \theta \approx 0.86\ x_Z \cos \theta,
\end{align}
that is proportional to the line-of-sight component of the magnetic field: $B_{\mathrm{los}} \propto x_Z \cos \theta$. To derive Eq.~(\ref{eq:pV}), we neglected the second-order contribution of the Zeeman effect to the total intensity profile, while we used $\mathrm{max}(\bar{\phi}')=-\mathrm{min}(\bar{\phi}')=\sqrt{2/\pi e}$ for the extremes of the circular polarization profile.

\begin{figure}[ht!]
\centering
\includegraphics[width=0.45\textwidth]{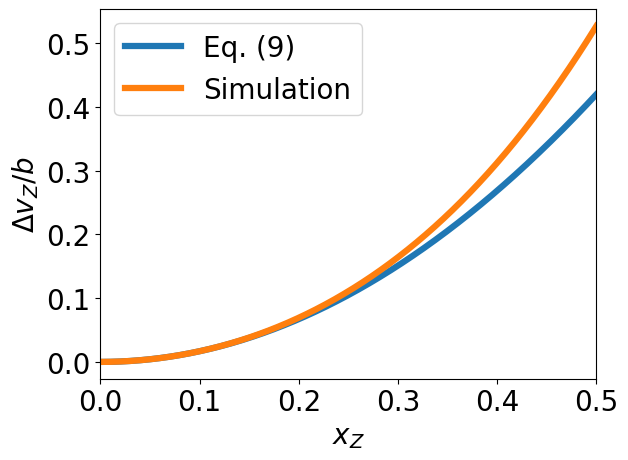}
\caption{Zeeman broadening of a Zeeman split line as a function of the Doppler normalized Zeeman shift $x_Z$. We compare the predicted Zeeman broadening between proper polarized radiative transfer simulations and the fitted Eq.~(\ref{eq:zeeman_broad}). The radiative transfer equations are performed using $\bar{Q}=-\Delta Q=1$ and $\cos \theta =1$.}. 
\label{fig:broaden_fit}
\end{figure}

From Eq.~(\ref{eq:zeeman_thin}a), we can see that the Zeeman effect also broadens spectral lines. We characterize the Zeeman broadening using the parameter,
\begin{align}
\Delta v_Z = \Delta v - \Delta v_{\mathrm{unsplit}},
\label{eq:zeeman_broad_parameter}
\end{align}
which is the difference between the full width at half maximum (FWHM) of the Zeeman split line $\Delta v$ and the unsplit line $\Delta v_{\mathrm{unsplit}}$. We find that the Zeeman broadening of the total intensity profile is well described by
\begin{align}
\label{eq:zeeman_broad}
\frac{\Delta v_Z}{b} \approx 0.84 x_Z^2 \left[ \bar{Q} - \Delta Q \cos^2 \theta  \right],
\end{align}
which, as can be seen from Fig.~(\ref{fig:broaden_fit}), is an excellent approximation, for $x_Z \lesssim 0.2$. Note here that, unlike the circular polarization which is only dependent on the line-of-sight component of the magnetic field, the Zeeman broadening is the sum of two components: one proportional to the total magnetic field strength and one proportional to the line-of-sight component of the magnetic field. These are the $x_Z^2$ ($\propto B^2$) and $x_Z^2 \cos^2 \theta$ ($\propto B_{\mathrm{los}}^2$) terms, respectively. 

The production of linear polarization through the Zeeman effect is a function of the plane-of-the-sky component of the magnetic field. From Eqs.~(\ref{eq:zeeman_thin}b,c) we see that the linear polarization fraction as a function of the magnetic field position angle, $Q_{\nu}' = \cos \left(2 \eta\right) Q_{\nu}+\sin \left(2\eta\right) U_{\nu}$, is given by
\begin{align}
\label{eq:pQ_fac}
p_l &= \frac{\left[ Q_{\nu}'\right]_{\mathrm{max}}}{I_{\mathrm{max}}} \simeq -\frac{1}{2} \ x_Z^2 \Delta Q \sin^2 \theta,  
\end{align}
where we note that the linear polarization fraction exhibits a quadratic dependence on the plane-of-the-sky magnetic field strength: $B_{\mathrm{pos}}^2 \propto x_Z^2 \sin^2 \theta$. The linear polarization has a spectral profile proportional to the second derivative of the total intensity, which is negative towards the line center and positive in the line wings. Therefore, for positive (negative) $\Delta Q$, the linear polarization is oriented perpendicular (parallel) to the projected magnetic field direction towards the line center, and parallel (perpendicular) to the projected magnetic field direction in the line wings. 

\subsubsection{Simultaneous observation of Zeeman signatures}
In Table~\ref{tab:B_sensitivity} we report the sensitivity of the different spectral line Zeeman signatures to the magnetic field components of the traced region. While circular polarization is only dependent on the line-of-sight magnetic field strength, the Zeeman broadening has also a dependence on the total magnetic field strength. When either of these methods are combined with linear polarization observations which are sensitive to (the square of) the two magnetic field components projected onto the plane-of-the-sky, then the full 3D vector of the magnetic field of the traced region can be readily derived. The circular polarization is linearly dependent on the (line-of-sight) magnetic field strength, while both Zeeman broadening and linear polarization observations are quadratically dependent on the magnetic field strength and are therefore insensitive to the polarity of the magnetic field vector. 
\begin{table}[ht!]
    \caption{Zeeman signatures and the magnetic field components they are sensitive to. The magnetic field components are divided up into their line-of-sight component, $B_{\mathrm{los}}$ and their two components in the plane-of-the-sky, $B_{x_\mathrm{pos}}$ and $B_{y_\mathrm{pos}}$.}
\label{tab:B_sensitivity}
    \centering
    \begin{tabular}{l c c}
    \hline \hline
    Technique & Sensitivity & Equation  \\
    \hline
Circular pol. & $B_{\mathrm{los}}$ & Eq.~(\ref{eq:pV}) \\ 
Zeeman broad. & $B^2$, $B_{\mathrm{los}}^2$ & Eq.~(\ref{eq:zeeman_broad}) \\ 
Linear pol. & $B_{x_\mathrm{pos}}^2 - B_{y_\mathrm{pos}}^2$, $B_{x_\mathrm{pos}}B_{y_\mathrm{pos}}$  & Eq.~(\ref{eq:pQ_fac}) \\ 
\hline
GK effect &  $\frac{B_{x_\mathrm{pos}}^2- B_{y_\mathrm{pos}}^2}{B^2}$, $\frac{B_{x_\mathrm{pos}}B_{y_\mathrm{pos}}}{B^2}$ & \\
Dust polarization &  $\frac{B_{x_\mathrm{pos}}^2- B_{y_\mathrm{pos}}^2}{B^2}$, $\frac{B_{x_\mathrm{pos}}B_{y_\mathrm{pos}}}{B^2}$ & \\
\hline
    \end{tabular}
\end{table}

Comparing the Zeeman signatures to methods such as observation of spectral line polarization through the GK effect or through dust polarization observations, we note the advantageous property of Zeeman signatures: their effect is a function of the magnetic field strength and not just its direction. Therefore, while methods using dust polarization or the GK effect may only trace the magnetic field direction, Zeeman signatures can be used to extract both the magnetic field direction and strength.

\subsubsection{Zeeman parameters}
To complete our discussion of Eq.~(\ref{eq:zeeman_thin}), we consider the Zeeman parameters in more detail. Each spectral line has associated with it three Zeeman parameters, the Zeeman coefficient, $z$, which is of the order kHz$/$mG, and the dimensionless coefficients $\bar{Q}$ and $\Delta Q$. The Zeeman coefficient, $z$, describes the relative offset of the $\sigma^{\pm}$-transition groups per unit magnetic field strength. Transitions with large Zeeman coefficients have relatively strong circular polarization associated with them. The dimensionless Zeeman coefficients $\bar{Q}$ and $\Delta Q$ describe the intragroup broadening of the  $\pi^0$ and $\sigma^{\pm}$ transitions. For a $J=1-0$ transition, the transition groups $\sigma^{\pm}$ and $\pi^0$ are each associated with only one transition. Therefore, for $J=1-0$ transitions, no additional broadening or linear polarization is produced due to intragroup broadening and the Zeeman broadening and linear polarization are perfectly described by the Zeeman coefficient: $\bar{Q}=-\Delta Q=1$ \citep[see also][]{crutcher:93}. However, for transitions of larger angular momentum, a multitude of transitions make up the $\sigma^{\pm}$ and $\pi^0$ transition groups, and they may exhibit intragroup broadening. The intragroup broadening adds to the total broadening of the spectral line, as well as to its linear polarization. The parameters $\bar{Q}$ and $\Delta Q$ are multiplication factors that address the relative contribution of intragroup broadening to the total Zeeman broadening or linear polarization. They are large if the intragroup broadening exceeds the relative offset of the $\sigma^{\pm}$-transition groups, whose effect is already captured by the Zeeman coefficient.
\begin{figure}[h!]
\centering
\includegraphics[width=0.49\textwidth]{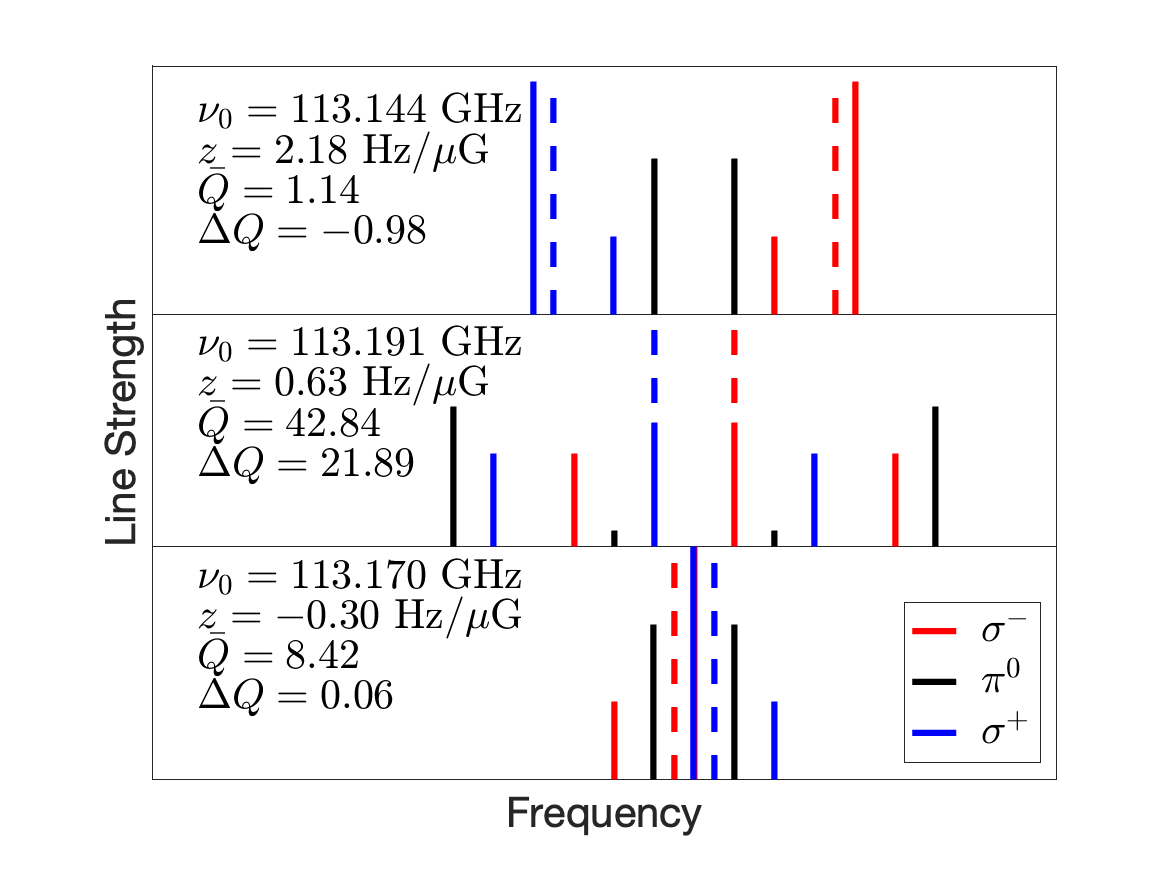}
\caption{Stem plots of the individual magnetic sublevel transitions of Band $3$ CN hyperfine transitions. Individual magnetic sublevel transitions are indicated by a solid line, and grouped ($\pi^0,\ \sigma^{\pm}$) by color. The dotted lines indicate the average shift of the $\sigma^{\pm}$ groups. }
\label{fig:zeeman_band3}
\end{figure}

In anticipation of the simulations that are reported in Section \ref{sec:sims}, it will be helpful to consider the spectral decomposition of Zeeman split lines from the CN Band 3 transitions. In Fig.~(\ref{fig:zeeman_band3}), we compare the Zeeman split transitions of the $\nu_0=113.144$ GHz, $\nu_0=113.191$ GHz, and $\nu_0=113.170$ GHz lines. These lines have different Zeeman effects and signatures. For the $\nu_0=113.144$ line, $\sigma^{\pm}$-transitions are shifted in frequency with respect to each other relatively strongly. This is why this transition exhibits strong circular polarization, reflected in its high $z$-parameter, while also having modest broadening or linear polarization, reflected in the $\bar{Q}$ and $\Delta Q$ parameters approaching $1$ and $-1$. The $\nu_0=113.191$ GHz line has a large spread in its Zeeman split transitions, even though the relative shift of the $\sigma^{\pm}$-transitions is modest. This transition exhibits therefore relatively low circular polarization (low $z$-parameter), but strong broadening and linear polarization (high $\bar{Q}$ and $\Delta Q$ parameters). The line at $\nu_0=113.170$ GHz shows generally weak Zeeman shifts in its magnetic sublevel transitions, which is why polarization and broadening is relatively weak for this line, reflected in its low $z$-, $\bar{Q}$ and $\Delta Q$ parameters. 


\subsection{Zeeman effects for molecules excited in protoplanetary disks}
In Section \ref{sec:sims}, we investigate the signature of the protoplanetary disk magnetic field in the spectral lines of CN. There, we use proper 3D polarized radiative transfer simulations to study Zeeman broadening, linear polarization and the emergence of circular polarization in a range of CN transitions. In anticipation of the discussion of these results, it will be helpful to analyze some simple radiative transfer problems for optically thin, Zeeman split lines that will help in interpreting the more intricate 3D simulations. 

Protoplanetary disks are permeated by a magnetic field which is commonly divided up into its cylindrical components: radial, vertical and toroidal components. It is a reasonable approximation to assume that these components are axisymmetric \citep{bethune:17}. The toroidal magnetic field component is likely dominant and, together with the radial magnetic field component, it changes sign between either sides of the disk \citep[][for more discussion, see text after Eq.~\ref{eq:B_field}]{bai:13,lesur:21}. 

Using the Zeeman effect, the magnetic field of protoplanetary disks is most effectively traced by the paramagnetic molecule CN. This is because CN exhibits a strong Zeeman effect and shows strong emission from an elevated emission surface at both sides of the disk \citep{cazzoletti:18}. We construct a simplified representation of the radiative transfer in a protoplanetary disk by dividing it up into two propagations through optically thin isothermal slabs which correspond to the back and front side molecular emission surfaces of the disk. We account only for the toroidal and vertical components of the magnetic field as the radial component is assumed to be negligible.

We consider a disk with inclination $\iota$ and position angle $\Phi_{\mathrm{PA}}$. The position angle is defined by the position angle of the disk redshifted semi-major axis. In the disk frame, where the $x$-axis is along its semi-major axis, and the $z$-axis is the rotation axis of the disk, the line-of-sight direction towards the observer, $\hat{n}_{\mathrm{los}}$, and the Northern and East direction in the plane-of-the-sky, $\hat{x}_{\mathrm{POS}}$ and $\hat{y}_{\mathrm{POS}}$, are
\begin{subequations}
\begin{align}
\hat{n}_{\mathrm{los}} &= \begin{pmatrix} -\sin\iota\sin \Phi_{\mathrm{PA}} \\ \sin\iota\cos \Phi_{\mathrm{PA}} \\ \cos\iota\end{pmatrix}, \\ \quad \hat{x}_{\mathrm{POS}} &= \begin{pmatrix} \cos \Phi_{\mathrm{PA}} \\ \sin \Phi_{\mathrm{PA}} \\ 0 \end{pmatrix} , \\ \hat{y}_{\mathrm{POS}} &= \begin{pmatrix} -\cos\iota\sin \Phi_{\mathrm{PA}} \\ \cos \iota \cos \Phi_{\mathrm{PA}} \\ -\sin \iota \end{pmatrix}.
\end{align}
\end{subequations}
From these definitions, we may compute the projection of the magnetic field onto the line-of-sight. We consider an axisymmetric magnetic field with a toroidal and a vertical component: $\boldsymbol{B}(r,\phi) = B_{\mathrm{v}}(r_c) \hat{\boldsymbol{z}} + B_{\mathrm{t}}(r_c) \hat{\boldsymbol{\phi}}$. The magnetic field strengths are dependent on the (disk-frame) cylindrical radius, $r_c$, and the unit vectors in the disk-frame are $\hat{\boldsymbol{z}}=[0,0,1]$ and $\hat{\boldsymbol{\phi}}=[-\sin \phi,\cos \phi,0]$ which are a function of the (disk-frame) azimuthal angle $\phi$. The projection of the magnetic field onto the line-of-sight then is,
\begin{align}
\label{eq:proj_simple}
\cos \theta = \hat{n}_{\mathrm{los}} \cdot \frac{\boldsymbol{B}}{B} = \frac{B_\mathrm{v}}{B} \cos \iota \pm \frac{B_\mathrm{t}}{B} \sin \iota \cos (\phi-\Phi_{\mathrm{PA}}), 
\end{align}
where $\pm$ is positive for the front and negative for the back side of the disk. Note that we assumed the toroidal component to be counter to, and along, the disk rotation direction for the front and back side of the disk, respectively. We suppressed the magnetic field strength dependence on the cylindrical radius in our notation. From now on, we note the azimuthal angle, $\phi' = \phi-\Phi_{\mathrm{PA}}$, with respect to the disk position angle. The magnetic field position angle, $\eta$, that is relevant to evaluate the linear polarization Stokes parameters may be evaluated from
\begin{align}
\label{eq:proj_eta}
\sin \theta\ e^{i\eta} &= \hat{x}_{\mathrm{POS}} \cdot \frac{\boldsymbol{B}}{B} + i\  \hat{y}_{\mathrm{POS}} \cdot \frac{\boldsymbol{B}}{B} \nonumber \\
&= \mp \frac{B_\mathrm{t}}{B} \sin \phi' + i \left( \pm \frac{B_{\mathrm{t}}}{B} \cos \iota \cos \phi'  - \frac{B_\mathrm{v}}{B} \sin \iota \right),
\end{align}
where the $\pm$ and $\mp$, apply for the front side and back side of the disk, respectively.

\subsubsection{Circular polarization}
We put together Eqs.~(\ref{eq:pV}) and (\ref{eq:proj_simple}) to obtain the expected circular polarization fraction, emerging from the front and back side of the disk,
\begin{align}
\label{eq:pV2}
p_V = 0.86 x_Z \left(\frac{B_\mathrm{v}}{B} \cos \iota \pm \frac{B_\mathrm{t}}{B} \sin \iota \cos \phi' \right),
\end{align}
where the $\pm$-sign corresponds to the front and back side of the disk. The contributions of the vertical and toroidal magnetic field components to the total circular polarization signal are additive, $p_V = p_V^{\mathrm{v}} + p_V^{\mathrm{t}}$, so we may consider them separately.
\begin{figure}[ht!]
\centering
\includegraphics[width=0.45\textwidth]{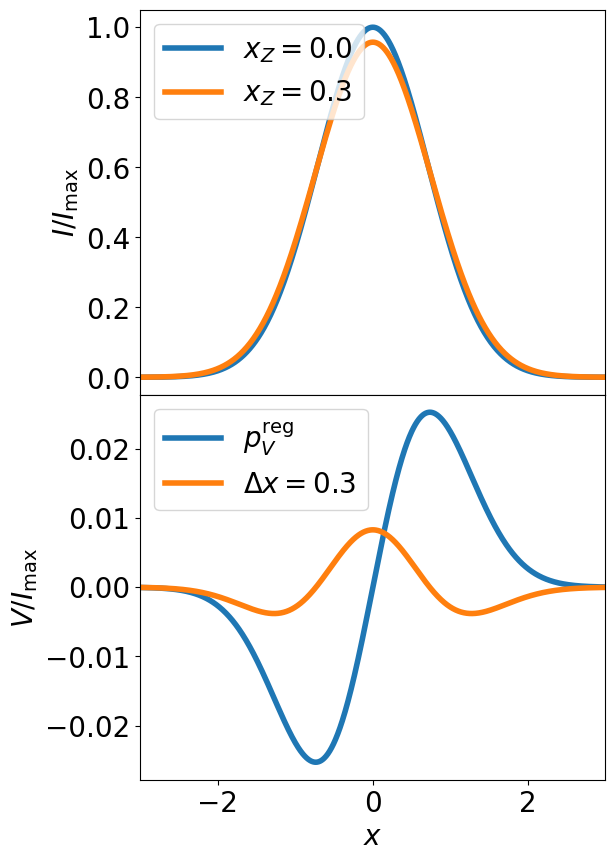} 
\caption{Total (a) and circularly polarized (b) intensity of a Zeeman split line. The intensities are normalized with respect to the unsplit ($x_Z=0$) intensity. The total intensity plots are for an unsplit ($x_Z=0$) and Zeeman split ($x_Z=0.3$, $\bar{Q}=-\Delta Q=1$, $\theta = \pi/2 - \pi/30$) line. The circularly polarized intensity plots are for a single propagation with constant magnetic field, $x_Z=0.3$ and $\theta = \pi/2 - \pi/30$, that yield the $p_V^{\mathrm{reg}}$ curve. The curve indicated by $\Delta x=0.3$ is the circular polarization from two propagations with a velocity shift $\Delta x = 0.3$ and oppositely oriented magnetic fields ($x_Z=0.3$ and $\theta = \pm \pi/2 - \pi/30$).} 
\label{fig:Stokes_profiles}
\end{figure}

The vertical component of the magnetic field does not change sign across the disk midplane and the vertical projection does not depend on the azimuthal angle. The contribution of the vertical magnetic field to the circular polarization is thus $p_V^{\mathrm{v}} = 0.86 x_Z B_{\mathrm{v}}/B \cos i$ for either sides of the disk. 

The toroidal magnetic field component changes sign through the midplane. Effectively, the circular polarization that is produced in the back side of the disk is then cancelled by the circular polarization produced in the front side of the disk. A net circular polarization is produced if, (i) a velocity shift along the line-of-sight occurs between the front and back side of the disk, (ii) the projected magnetic field varies between both sides of the disk, or (iii) an optically thick dust layer is present between both emission layers that partially absorbs the emission emerging from the back side of the disk \citep[as could be seen in, e.g.,~][]{isella:18}. 

The recovery of circular polarization through a velocity shift between the emission surfaces and a difference in the projected magnetic fields of both emission surfaces is proportional to the inclination. Of these two effects, we found that a velocity shift between the emission surfaces is most effective in negating the circular polarization cancellation. The velocity shift along the line-of-sight between the emission surfaces occurs for inclined disks with emission surfaces above the midplane. It can be larger than the line-width at large inclinations $\iota \gtrsim 40^o$, and roughly follows an azimuthal profile $\propto \sin 2 \phi'$. In large parts of the disk two well-separated line profiles would emerge that do not interfere in each others (polarized) emission. For moderately inclined disks, $\iota \lesssim 20^o$, the velocity shifts are generally smaller than the Doppler widths for all azimuthal angles. We may evaluate the impact of a velocity shift on the emergent circularly polarized emission. We let $\Delta x = (v_{\mathrm{upper}} - v_{\mathrm{lower}})/b$ denote the (Doppler normalized) velocity shift between the line-of-sight projected velocities of the emission emerging from the front and back side of the disk. For $\Delta x \lesssim 0.3$, the two-step (optically thin) radiative transfer with opposite magnetic fields permits the solution for the circular polarization (see Appendix \ref{sec:ap_B} for a derivation)
\begin{align}
\label{eq:pV_vel}
V_{\nu} \simeq \frac{x_Z \Delta x }{2} \frac{B_{\mathrm{t}}}{B} \sin \iota \cos \phi' \frac{d^2 I_{\nu}}{dx^2},
\end{align}
that yields a polarization fraction of 
\begin{align}
p_V^{\mathrm{t,vel}} &\simeq \frac{1}{2}\left(\frac{2}{e^{3/2}} + 1 \right) x_Z \frac{B_{\mathrm{t}}}{B} \sin \iota \cos \phi' \ \Delta x \nonumber \\ &\approx 0.84\ \Delta x\ p_V^{\mathrm{t,reg}} ,
\end{align}
where $p_V^{\mathrm{t,reg}} = 0.86 x_Z \frac{B_\mathrm{t}}{B} \sin \iota \cos \phi'$ is the expected circular polarization fraction due to the toroidal magnetic field from the front side of the disk. Thus, some of the circular polarization due to the toroidal magnetic field that would have been cancelled through the magnetic field sign-flip is recovered when a velocity shift between both emission surfaces is present. Note, however, that the line profile of the circular polarization in this special case is not the usual $S$-shaped profile as expected from the `regular' Zeeman effect, but rather a profile proportional to $\frac{d^2 I_{\nu}}{dx^2}$, as shown in Fig.~(\ref{fig:line_profiles}).

Another way to suppress the destructive interference of the toroidal magnetic field circular polarization is through an optically thick dust layer in the midplane. If absorption due to dust occurs in between the back and front side emission surfaces then the cancellation is partially suppressed. In this scenario, the resulting circular polarization can be estimated as (see Appendix \ref{sec:ap_B} for a derivation)
\begin{align}
\label{eq:pV_dust}
p_V^{\mathrm{t}} \simeq p_V^{\mathrm{t,reg}} \tanh{\frac{\tau_{\mathrm{dust}}}{2}} + \frac{2p_V^{\mathrm{t,vel}}}{e^{\tau_{\mathrm{dust}}}+1} .
\end{align}
As an example, we estimate the effect of a dust layer on the circular polarization in a source such as TW Hya. Estimates of TW Hya's dust optical depth at $50$ AU for a range of frequencies put these at $\tau_{\mathrm{dust}}=0.7$ to $2$ from ALMA's Band $3$ to Band $7$ \citep{macias:21}. That means that $33-76\%$ (Band $3$ $-$ Band $7$) of the $p_V^{\mathrm{t,reg}}$ is recovered when one accounts for the dust absorption. 

\subsubsection{Zeeman broadening}
We combine Eqs.~(\ref{eq:zeeman_broad}) and (\ref{eq:proj_simple}), to obtain the expected (Doppler normalized) Zeeman broadening emerging from the front and back side of the disk,
\begin{align}
\label{eq:broad_disk}
\frac{\Delta v_Z}{b} &= 0.84 x_Z^2 \left[\bar{Q} - \Delta Q \left(\frac{B_{\mathrm{v}}^2}{B^2} \cos^2 \iota + \frac{B_{\mathrm{t}}^2}{B^2} \sin^2 \iota \right. \right. \nonumber \\ &\times \left. \left.\cos^2 \phi'  \pm \frac{B_{\mathrm{v}} B_{\mathrm{t}}}{B^2} \sin 2i \sin \phi' \right) \right],
\end{align}
where the $\pm$-sign corresponds to the front and back side of the disk. We recognize from Eq.~(\ref{eq:broad_disk}) that most of the Zeeman broadening is insensitive to the polarity of the magnetic field and comes to expression in the same way in both the front and back side of the disk. Only the cross-term $B_{\mathrm{v}}B_{\mathrm{t}}$, is explicitly dependent on the polarity of the toroidal magnetic field which changes sign through the midplane. For weakly inclined disks the cross-term is minor compared to broadening terms that are invariant to the magnetic field polarity. The expression of the cross-term is therefore dependent on the presence of a dust layer and on the velocity shift between the front and back side of the disk emission surfaces. 


We previously pointed out that when considering inclined disks, a velocity shift arises between the back and front side of the disk. For moderately inclined disks this velocity shift manifests in the total intensity profile of spectral lines as an additional broadening term. For a Doppler normalized velocity shift between both emission surfaces of $\Delta x$, we note the additional broadening
\begin{align}
\frac{\Delta v_{\mathrm{shift}}}{b} \simeq 0.84 \left(\Delta x\right)^2,
\end{align}
where the total broadening of a Zeeman split line for a slightly inclined disk is the sum of both broadening mechanisms, $\Delta v_Z + \Delta v_{\mathrm{shift}}$, if they are small compared to the line width. The Zeeman broadening can be disentangled from the broadening due to the velocity shift between both emission surfaces by observing multiple spectral lines. While the $\Delta v_Z$ contribution to the broadening is dependent on the Zeeman coefficient which varies between transitions, the $\Delta v_{\mathrm{shift}}$ contribution to the broadening is dependent on the system geometry and does not vary between transitions. The presence of an optically thick dust layer in the midplane will suppress the broadening due to the velocity shift between both emission surfaces.

\subsubsection{Linear Polarization}
The linear polarization of Zeeman split spectral lines is quadratically dependent on the plane-of-the-sky component of the magnetic field [see also, Eqs.~(\ref{eq:zeeman_thin}b,c) and (\ref{eq:proj_eta})]. Linear polarization is observed through the Stokes $Q$ and $U$ parameters: the simultaneous observation of these parameters yield a two-dimensional quasi vector, which may be related to the magnetic field direction (but not its polarity) and strength projected onto the plane of the sky. We put together Eqs.~(\ref{eq:zeeman_thin}b,c) and (\ref{eq:proj_eta}) to derive an expression for the linear polarization fractions, $q_{\nu_0} = Q_{\nu_0}/I_{\nu_0}$ and $u_{\nu_0} = U_{\nu_0}/I_{\nu_0}$, at the line-center, $\nu_0$,
\begin{subequations}
\label{eq:qu_fracs}
\begin{align}
q_{\nu_0} &= -\frac{x_Z^2 \Delta Q}{2} \left(-\frac{B_{\mathrm{v}}^2}{B^2} \sin^2 \iota  - \frac{B_{\mathrm{t}}^2}{B^2}  \left[\cos 2\phi' - \sin^2 \iota \right. \right. \nonumber \\  &\times \left. \left. \cos^2 \phi' \right] \pm  \frac{B_\mathrm{v}B_\mathrm{t}}{B^2} \sin 2\iota \cos \phi' \right) \\
u_{\nu_0} &= -\frac{x_Z^2 \Delta Q}{2} \left( -\frac{B_{\mathrm{t}}^2}{B^2}\cos \iota \sin 2\phi' \pm 2\frac{B_\mathrm{v}B_\mathrm{t}}{B^2} \sin \iota \sin \phi' \right),
\end{align}
while the total linear polarization fraction at the line-center, $p_l = \sqrt{q_{\nu_0}^2+u_{\nu_0}^2}$, is 
\begin{align}
p_l &= -\frac{x_Z^2 \Delta Q}{2} \left(\frac{B_{\mathrm{t}}^2}{B^2}\left[\sin^2 \phi' + \cos^2 \iota \cos^2 \phi' \right] + \frac{B_{\mathrm{v}}^2}{B^2} \sin^2 \iota \right. \nonumber \\  &\mp \left. \frac{B_{\mathrm{t}} B_{\mathrm{v}}}{B^2} \sin 2\iota \cos \phi' \right).
\end{align}
\end{subequations}
The $\pm$-sign applies for the front and back side of the disk. Just as for the Zeeman broadening, a polarity-dependent term arises in the expressions for the linear polarization. The manifestation of the cross-term is dependent on the presence of a dust layer and on the velocity shift between the front and back side of the disk emission surfaces, and is maximal at inclinations of $45^o$. 

\subsubsection{Azimuthal stacking of Zeeman signatures}
We have seen that the Zeeman broadening and linear polarization are partially dependent on terms that change sign between the back and front side of the disk. The polarity-dependent terms are maximal at a disk inclination of $45^o$, and require detailed knowledge of the continuum absorption or a velocity shift between the two emission surfaces along the line-of-sight in order for them to be modeled properly. In this way, the interpretation of the Zeeman broadening or linear polarization measurements for their magnetic field information is, just like circular polarization observations, inherently associated with uncertainty. In order to remove this uncertainty, we take the axisymmetry of the disk magnetic field into consideration such that we may devise an averaging scheme which eliminates the polarity dependent term, similar to the technique employed in \citet{teague:21}. Namely, when one stacks the Zeeman broadening over the range of deprojected azimuthal angles, 
\begin{align}
\label{eq:broad_average}
&\Braket{\frac{\Delta v_Z}{b}}_{\phi'} = 0.84 x_Z^2 \left[\bar{Q} - \Delta Q 
\Braket{\cos^2 \theta}_{\phi'}  \right] \nonumber \\
&= 0.84 x_Z^2 \left[\bar{Q} - \frac{\Delta Q}{2} \left(\frac{B_{\mathrm{v}}^2}{B^2}\cos^2 \iota + \frac{B_{\mathrm{t}}^2}{B^2} \sin^2 \iota ] \right)\right],
\end{align}
then the resulting averaged quantity is insensitive to the polarity-dependent terms in $\cos^2 \theta$. Similarly, we may exploit the axisymmetry of the disk to extract the different magnetic field components from the linear polarization by applying the weighted stacking,
\begin{subequations}
\begin{align}
\braket{q_{\nu} \cos 2\phi' + u_{\nu} \sin 2\phi'}_{\phi'}&= \frac{x_Z^2 \Delta Q}{2} \frac{B_\mathrm{t}^2}{B^2} \nonumber \\ &\times \frac{1+\cos \iota - \frac{1}{2}\sin^2 i}{2},
\end{align}
to extract the toroidal magnetic field component, and the weighting
\begin{align}
\braket{q_{\nu}}_{\phi'} = \frac{x_Z^2 \Delta Q}{2} \sin^2 \iota \left(B_{\mathrm{v}}^2 - \frac{B_{\mathrm{t}}^2}{2}\right),
\end{align}
\end{subequations}
to obtain the vertical magnetic field component.

\section{Simulations}
\label{sec:sims}
In this section, we study the signature of magnetic fields in CN lines excited in a TW Hya like protoplanetary disk using full 3D radiative transfer simulations. We model the polarized radiative transfer using Eqs.~(\ref{eq:line})-(\ref{eq:pol_abs}), which are based on Equations (9.2-9.16) from \citet{landi:06}. In this formalism, one rigorously models the propagation of radiation through a Zeeman split population by resolving the individual magnetic sublevel transitions within a line. A similar method is employed in the radiative transfer code \texttt{POLARIS} \citep{brauer:17}.

\begin{table*}[]
    \caption{Summary of the simulation results of the broadening and circular polarization of the $N=1-0$ transitions of CN at deprojected distance $r_c=50$ AU and $\phi=0^o$ and $\phi = 90^o$. Results of simulations assuming CN emission surfaces at $z/r=0.1$ with abundances $x_{\mathrm{CN}}=1\times10^{-8},\ 1\times10^{-9}, \ \mathrm{and} \ 1\times10^{-10}$, and $z/r=0.3$ with abundances $x_{\mathrm{CN}}=3\times10^{-7},\ 3\times10^{-8}, \ \mathrm{and} \ 3\times10^{-9}$ are reported.}
\label{tab:rc50_sims}
    \centering
    \begin{tabular}{l l c c c c c c}
    \hline \hline
     & &  & $z/r=0.1$ &  & & $z/r=0.3$ &   \\
    & $x_{\mathrm{CN}}$= & $10^{-10}$ & $10^{-9}$ & $10^{-8}$ & $3\times10^{-9}$ & $3\times10^{-8}$ & $3\times10^{-7}$ \\
    \hline
 $\nu_0 = 113.14416$ GHz 
 &  $\tau_{CN}$ & $0.08$ &  $0.84$ &  $8.46$ &  $0.08$ &  $0.73$ &  $6.65$ \\ 
 &  $\Delta v_Z$ (m/s) & $39.94$ &  $40.29$ &  $47.40$ &  $27.32$ &  $27.49$ &  $34.26$ \\ 
 &  $p_l$ (\%) & $14.41$ &  $10.31$ &  $0.09$ &  $7.05$ &  $5.11$ &  $0.08$ \\ 
 &  $p_V^{\mathrm{dust}}$ (\%) & $2.49$ &  $2.49$ &  $3.07$ &  $2.03$ &  $2.05$ &  $2.49$ \\ 
 &  $p_V^{\mathrm{thin}}$ (\%) & $1.42$ &  $1.34$ &  $1.58$ &  $1.02$ &  $0.97$ &  $1.15$ \\ 
   $\nu_0 = 113.12337$ GHz 
 &  $\tau_{CN}$ & $0.01$ &  $0.10$ &  $1.04$ &  $0.01$ &  $0.08$ &  $0.79$ \\ 
 &  $\Delta v_Z$ (m/s) & $14.25$ &  $14.28$ &  $14.40$ &  $9.46$ &  $9.66$ &  $9.74$ \\ 
 &  $p_l$ (\%) & $3.88$ &  $3.72$ &  $2.30$ &  $1.86$ &  $1.80$ &  $1.23$ \\ 
 &  $p_V^{\mathrm{dust}}$ (\%) & $0.80$ &  $0.80$ &  $0.79$ &  $0.62$ &  $0.62$ &  $0.62$ \\ 
 &  $p_V^{\mathrm{thin}}$ (\%) & $0.46$ &  $0.45$ &  $0.42$ &  $0.31$ &  $0.31$ &  $0.29$ \\ 
   $\nu_0 = 113.19128$ GHz 
 &  $\tau_{CN}$ & $0.10$ &  $1.07$ &  $10.76$ &  $0.10$ &  $0.94$ &  $8.48$ \\ 
 &  $\Delta v_Z$ (m/s) & $148.85$ &  $142.04$ &  $133.90$ &  $96.08$ &  $94.36$ &  $104.36$ \\ 
 &  $p_l$ (\%) & $26.64$ &  $19.89$ &  $0.16$ &  $13.03$ &  $9.22$ &  $0.10$ \\ 
 &  $p_V^{\mathrm{dust}}$ (\%) & $0.78$ &  $0.74$ &  $0.64$ &  $0.61$ &  $0.60$ &  $0.61$ \\ 
 &  $p_V^{\mathrm{thin}}$ (\%) & $0.45$ &  $0.39$ &  $0.29$ &  $0.30$ &  $0.28$ &  $0.26$ \\ 
   $\nu_0 = 113.17049$ GHz 
 &  $\tau_{CN}$ & $0.08$ &  $0.83$ &  $8.28$ &  $0.07$ &  $0.68$ &  $6.29$ \\ 
 &  $\Delta v_Z$ (m/s) & $5.36$ &  $5.52$ &  $8.41$ &  $3.57$ &  $3.68$ &  $5.41$ \\ 
 &  $p_l$ (\%) & $0.06$ &  $0.04$ &  $0.00$ &  $0.02$ &  $0.01$ &  $0.00$ \\ 
 &  $p_V^{\mathrm{dust}}$ (\%) & $0.36$ &  $0.36$ &  $0.51$ &  $0.29$ &  $0.29$ &  $0.38$ \\ 
 &  $p_V^{\mathrm{thin}}$ (\%) & $0.20$ &  $0.20$ &  $0.27$ &  $0.14$ &  $0.14$ &  $0.18$ \\ 
   $\nu_0 = 113.49097$ GHz 
 &  $\tau_{CN}$ & $0.28$ &  $2.85$ &  $28.58$ &  $0.24$ &  $2.38$ &  $22.01$ \\ 
 &  $\Delta v_Z$ (m/s) & $9.03$ &  $10.36$ &  $18.20$ &  $6.42$ &  $7.00$ &  $12.38$ \\ 
 &  $p_l$ (\%) & $0.37$ &  $0.08$ &  $0.00$ &  $0.21$ &  $0.06$ &  $0.00$ \\ 
 &  $p_V^{\mathrm{dust}}$ (\%) & $0.65$ &  $0.73$ &  $1.16$ &  $0.53$ &  $0.58$ &  $0.89$ \\ 
 &  $p_V^{\mathrm{thin}}$ (\%) & $0.37$ &  $0.38$ &  $0.58$ &  $0.26$ &  $0.27$ &  $0.41$ \\ 
   $\nu_0 = 113.50891$ GHz 
 &  $\tau_{CN}$ & $0.08$ &  $0.83$ &  $8.31$ &  $0.08$ &  $0.74$ &  $6.60$ \\ 
 &  $\Delta v_Z$ (m/s) & $24.07$ &  $24.32$ &  $31.44$ &  $16.43$ &  $16.49$ &  $21.88$ \\ 
 &  $p_l$ (\%) & $6.98$ &  $4.89$ &  $0.03$ &  $3.39$ &  $2.42$ &  $0.03$ \\ 
 &  $p_V^{\mathrm{dust}}$ (\%) & $1.90$ &  $1.92$ &  $2.38$ &  $1.53$ &  $1.54$ &  $1.91$ \\ 
 &  $p_V^{\mathrm{thin}}$ (\%) & $1.09$ &  $1.03$ &  $1.21$ &  $0.77$ &  $0.73$ &  $0.89$ \\ 
   $\nu_0 = 113.48812$ GHz 
 &  $\tau_{CN}$ & $0.10$ &  $1.08$ &  $10.79$ &  $0.09$ &  $0.90$ &  $8.27$ \\ 
 &  $\Delta v_Z$ (m/s) & $76.92$ &  $74.86$ &  $85.08$ &  $49.64$ &  $49.46$ &  $62.57$ \\ 
 &  $p_l$ (\%) & $11.21$ &  $7.62$ &  $0.03$ &  $5.82$ &  $3.99$ &  $0.02$ \\ 
 &  $p_V^{\mathrm{dust}}$ (\%) & $2.42$ &  $2.39$ &  $2.82$ &  $2.01$ &  $2.01$ &  $2.40$ \\ 
 &  $p_V^{\mathrm{thin}}$ (\%) & $1.38$ &  $1.27$ &  $1.42$ &  $0.99$ &  $0.94$ &  $1.10$ \\ 
   $\nu_0 = 113.52043$ GHz 
 &  $\tau_{CN}$ & $0.01$ &  $0.10$ &  $1.04$ &  $0.01$ &  $0.10$ &  $0.87$ \\ 
 &  $\Delta v_Z$ (m/s) & $21.47$ &  $21.55$ &  $21.85$ &  $14.89$ &  $15.18$ &  $15.25$ \\ 
 &  $p_l$ (\%) & $7.26$ &  $6.96$ &  $4.38$ &  $3.59$ &  $3.43$ &  $2.28$ \\ 
 &  $p_V^{\mathrm{dust}}$ (\%) & $1.80$ &  $1.80$ &  $1.82$ &  $1.44$ &  $1.44$ &  $1.47$ \\ 
 &  $p_V^{\mathrm{thin}}$ (\%) & $1.04$ &  $1.03$ &  $0.97$ &  $0.74$ &  $0.73$ &  $0.69$ \\ 
   $\nu_0 = 113.49964$ GHz 
 &  $\tau_{CN}$ & $0.08$ &  $0.85$ &  $8.48$ &  $0.09$ &  $0.78$ &  $6.86$ \\ 
 &  $\Delta v_Z$ (m/s) & $50.11$ &  $50.53$ &  $54.67$ &  $33.63$ &  $33.86$ &  $40.80$ \\ 
 &  $p_l$ (\%) & $18.40$ &  $13.37$ &  $0.16$ &  $8.85$ &  $6.33$ &  $0.09$ \\ 
 &  $p_V^{\mathrm{dust}}$ (\%) & $0.90$ &  $0.87$ &  $0.75$ &  $0.65$ &  $0.64$ &  $0.68$ \\ 
 &  $p_V^{\mathrm{thin}}$ (\%) & $0.51$ &  $0.46$ &  $0.33$ &  $0.33$ &  $0.30$ &  $0.29$ \\ 
    \hline
    \end{tabular}
\end{table*}

We modeled the radiative transfer of CN in a TW Hya like disk, at inclination $\iota=6^o$, where we used the physical model of \citet{calahan:21}. In our models, we assume CN to be excited in an emission surface located at $z/r=0.1 \pm 0.025$ or $z/r=0.3 \pm 0.025$. Of these emission surface heights, $z/r \sim 0.3$ is predicted by chemical models \citep{cazzoletti:18}, while we include simulations with $z/r \sim 0.1$ in our analysis to analyze the sensitivity of our results to the emission height. According to the physical model we adopt, at cylindrical radius $r_c=50$ AU, emission from $z/r=0.3$ is associated with $50$ K gas, while emission from $z/r=0.1$ is associated with $22$ K gas. To cover a range of optical depths, corresponding roughly to $\tau \sim 0.1,\ 1 \ \mathrm{and}\ 10$, we adopted the emission surface abundances: $x_{\mathrm{CN}}=3\times10^{-7},\ 3\times10^{-8}, \ \mathrm{and} \ 3\times10^{-9}$ for the $z/r \sim 0.3$ simulations, and $x_{\mathrm{CN}}=1\times10^{-8},\ 1\times10^{-9}, \ \mathrm{and} \ 1\times10^{-10}$ for the $z/r \sim 0.1$ simulations. In order to evaluate the effects of dust absorption in the midplane, we performed simulations with and without a midplane layer of dust, with optical depth based on \citet{macias:21} and a vertical profile where the dust scale height is set to $20\%$ of the gas scale height. 

We adopted a magnetic field, based on the constraints of \citet{vlemmings:19},
\begin{align}
\label{eq:B_field}
B(\boldsymbol{r}_c,z) = \left(\frac{r_c}{50\ \mathrm{AU}} \right)^{-3/2} \left(25 \ \mathrm{mG} \  \mathrm{sgn}(z)  \hat{\boldsymbol{\phi}} + 0.8 \ \mathrm{mG} \  \hat{\boldsymbol{z}}\right),
\end{align}
with components in the toroidal, $\hat{\boldsymbol{\phi}}$, and vertical, $\hat{\boldsymbol{z}}$, direction. It is commonly assumed that the radial and toroidal component of magnetic fields that permeate accretion disks exhibit a sign-change crossing the midplane \citep{blandford:82, wardle:93, bai:16}. Indeed, such a magnetic field structure is expected from the advection of magnetic field lines throughout the star formation process. In the early evolutionary stages, field lines are dragged inwards \citep{mouschovias:99, girart:99}, creating a radial magnetic field of opposite polarity between both sides of the midplane. Subsequently, Keplerian rotation of the accretion disk winds the radial field lines into a toroidal magnetic field with opposite polarity at either side of the midplane. Accordingly, such a magnetic field configuration is required to magnetocentrifugally launch the often observed bipolar jets and winds associated with young stellar objects \citep{frank:14, bjerkeli:19}. Recent three-dimensional MHD simulations find that under specific conditions, the usual symmetry of the magnetic field can be broken, specifically for disks where rotation is anti-aligned with the magnetic field—such disks will then have an associated dissymmetrical wind \citep{bethune:17, bai:17}. Still, considering the magnetic field evolution and observational evidence in earlier stages of the star formation process, we adopt our simulations to the common expectation of a magnetic field that changes sign in its radial and toroidal component through the midplane.

The non-LTE excitation analysis was performed using LIME\footnote{We used LIME version 1.9.5, available from \url{https://github.com/lime-rt/lime}.} \citep{brinch:10}. The excitation solutions were subsequently used in conjunction with the magnetic field model to ray-trace a (polarized) image using the earlier mentioned formalism outlined in \citet{landi:06}\footnote{The source code for the polarized ray-tracing is available from \url{https://github.com/blankhaar/zeeman_disk}.}. We include in our analysis the 9 strongest Band $3$ transitions of CN which are part of a (hyper)fine manifold and therefore lie close in frequency around $113.4$ GHz. Due to the low inclination of the TW Hya like disk that we model, line overlap within the manifold does not occur, but this may be a possibility in more strongly inclined disks. The energy levels, Einstein coefficients and collision coefficients of CN were taken from the LAMDA database \citep{schoier:05, kalugina:15}. The level specific g-factors of CN were evaluated using the method outlined in Appendix A of \citet{vlemmings:19}. We analyze the (polarization) spectra and properties of the Band $3$ transitions of CN emerging at deprojected distance $r_c=50$ AU ($0.83"$). We choose to analyze the Band $3$ transitions as these exhibit the largest Doppler normalized Zeeman coefficients (see Table~\ref{tab:CN_mol} and Eq.~\ref{eq:zeeman_fo}). Around deprojected distance of $r_c=50$ AU CN emission is observed to peak \citep{cazzoletti:18}, while dust continuum remains important to the radiative transfer \citep{vlemmings:19, macias:21}. 

\begin{figure*}[ht!]
\centering
\includegraphics[width=\textwidth]{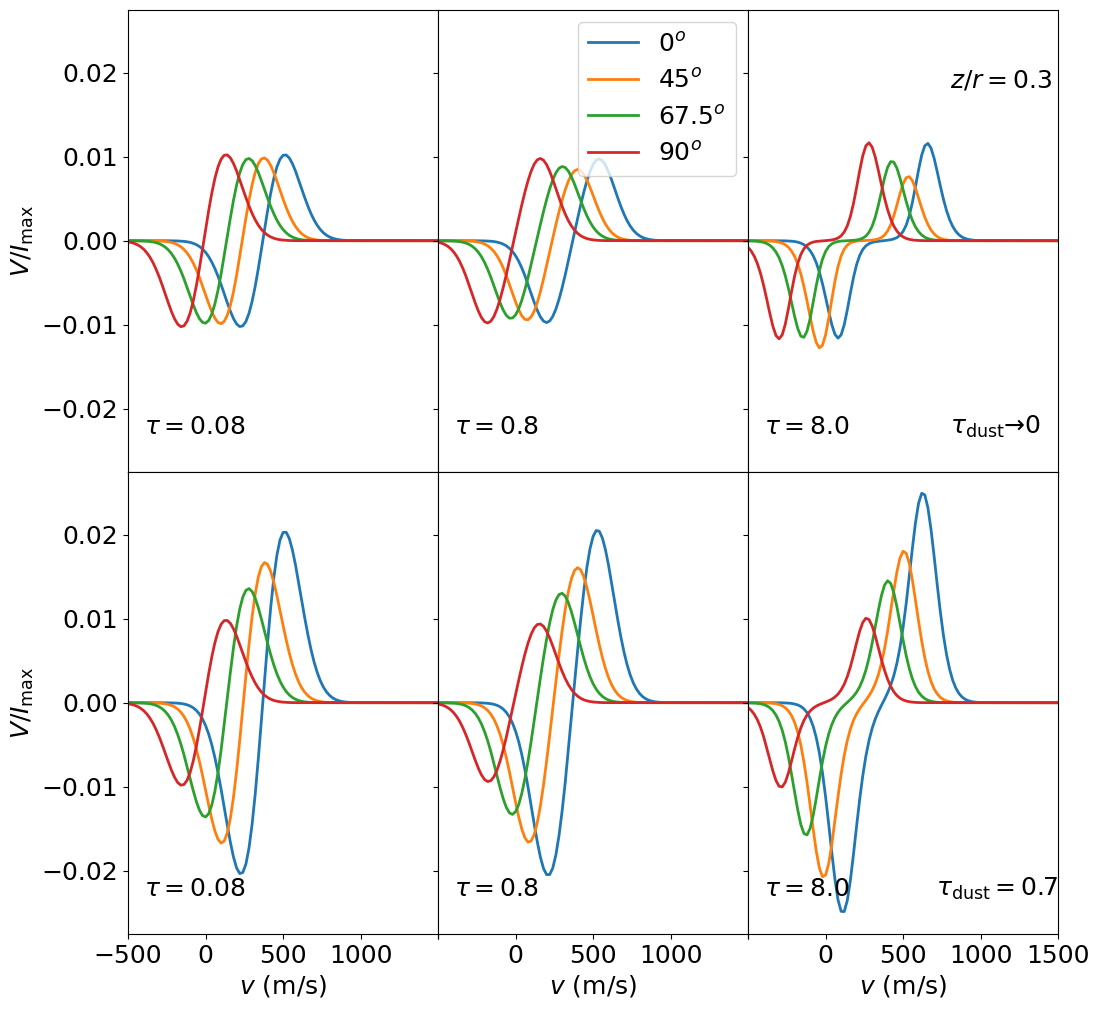}
\caption{Spectra of the circularly polarized intensity of the CN $(N,J,F) \to (N',J',F')=(1,1/2,1/2) \to (0,1/2,1/2)$ transition excited in a TW Hya like protoplanetary disk, in an emission surface around $z/r=0.3$, with (\textit{lower row}) and without (\textit{upper row}) a layer of dust present between the emission surfaces. The spectra for different optical depths and position angles are given.}
\label{fig:Stokes_V_50}
\end{figure*}

\subsection{Circular polarization}
The spectral profiles of the circular polarization of the CN $(N,J,F) \to (N',J',F')=(1,1/2,1/2) \to (0,1/2,1/2)$ transition at $r_c=50$ AU are shown in Fig.~(\ref{fig:Stokes_V_50}) at a range of azimuthal angles, $\phi'$. Simulations were performed with and without an optically thick midplane dust layer, and a large difference between both simulations can be readily appreciated. The simulations without a midplane dust-layer show $S$-shaped spectral profile in the circular polarization, with a circular polarization fraction of $\sim 1 \%$ that is weakly variable over the azimuthal angle. While also exhibiting $S$-shaped spectral profiles, the simulations that include an optically thick midplane dust-layer show a variation in the circular polarization fraction from $\sim 1 \% - 2 \%$ for emission emerging from different azimuthal angles. 

The variability of the circular polarization fraction with the azimuthal angle may be explained by the Zeeman effect due to the toroidal magnetic field. As discussed in Section 2.2.1, the circular polarization due to the toroidal magnetic field is affected by destructive interference between the emission from the front and back side of the disk. The interference is suppressed by either a velocity shift between both disk sides along the line of sight, or through an optically thick layer of dust. We note that that toroidal component comes to expression for the simulations with a midplane dust-layer, where we find the strongest circular polarization for the simulations with a dusty midplane at azimuthal angle $\phi'=0^o$ as the toroidal magnetic field contributes maximally to the line-of-sight component of the magnetic field.  In contrast, we find that the contribution to the circular polarization from the toroidal magnetic field component is almost completely suppressed for the simulations assuming optically thin dust. This may be explained by the small velocity shift between both emission surfaces: $\Delta x \approx 0.05$, due to the low inclination adopted. The small variation in the circular polarization that is a result of the velocity shift between the back and front side emission surface is most strongly present around azimuthal angles of $45^o$, as there the velocity shift is maximal. 

\begin{figure}[h]
\centering
\includegraphics[width=0.4\textwidth]{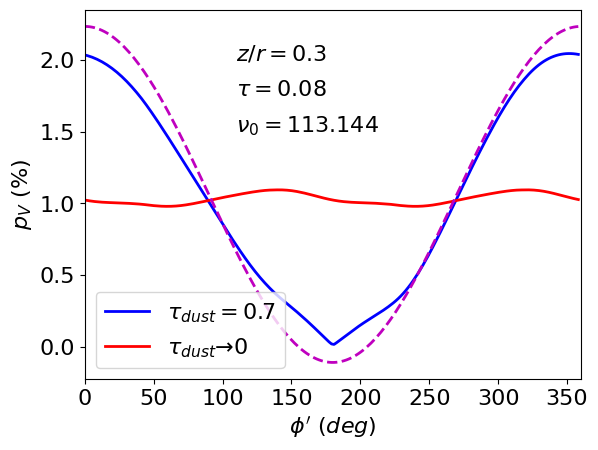}
\caption{Circular polarization fraction at deprojected distance $r_c=50$ AU, as a function of the deprojected azimuthal angle $\phi'$. Simulations with and without a dust-layer are plotted inside the figure and the dashed line is a fitting function.}
\label{fig:pV_psi}
\end{figure}

For the optically thin simulations with the emission surface at $z/r=0.3$, we have plotted the variability of the circular polarization fraction, defined in Eq.~(\ref{eq:circ_pol_parameter}), with the azimuthal angle in Fig.~(\ref{fig:pV_psi}). We note for the simulations with $\tau_{\mathrm{dust}}=0.7$ an enhanced polarization fraction for azimuthal angles $<180^o$, while it is diminished for azimuthal angles $>180^o$. Strikingly, the circular polarization that is predicted from our analytical modeling, $p_V = p_V^{\mathrm{v}} + p_V^{\mathrm{t}} \tanh{\frac{\tau_{\mathrm{dust}}}{2}}$ (see Eqs.~\ref{eq:pV2}-\ref{eq:pV_dust} and Section 2.2.1), compares well to the simulation results. The variation in the $\tau_{\mathrm{dust}}\to 0$ simulations is much weaker, due to the low velocity shift between both sides of the disk.

In Table~\ref{tab:rc50_sims}, we list the circular polarization of the different CN Band $3$ transitions we found in our simulations. The simulations show that the differently placed emission surfaces at $z/r=0.1$ and $z/r=0.3$ yield different circular polarization fractions for the same transition. This is the result of the different thermal line width for both surfaces, with the higher surface tracing warmer gas resulting in enhanced thermal line widths. Between transitions, we also find strong variations in the circular polarization. The circular polarization is weakly dependent on the optical depth when lines are $\lesssim 1$ in optical depth, and they increase slightly for $\tau \gtrsim 1$, as the gradient towards the line wings increase \citep[this effect was also seen in ][]{mazzei:20}.

\subsection{Zeeman broadening}
\begin{figure*}[ht!]
\centering
\includegraphics[width=\textwidth]{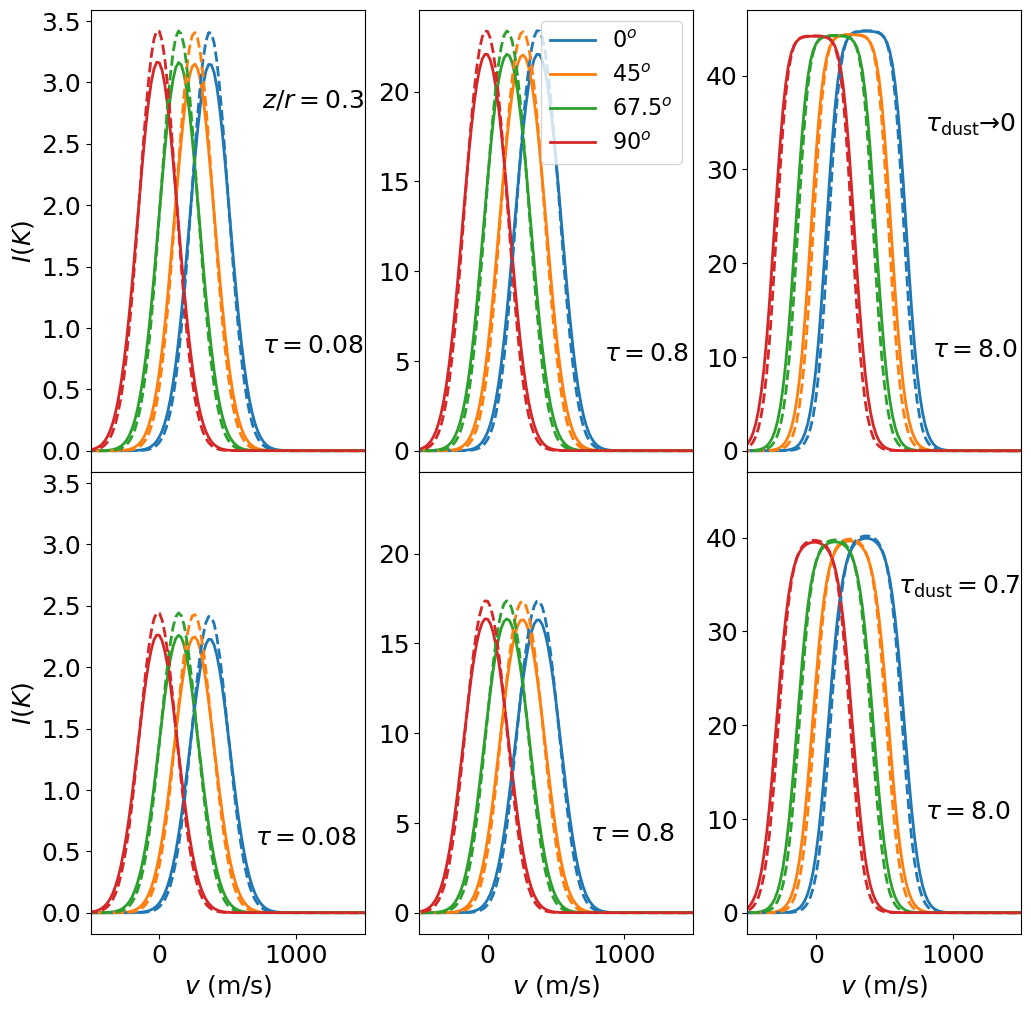}
\caption{Spectra of the total intensity of the CN $(N,J,F) \to (N',J',F')=(1,1/2,1/2) \to (0,1/2,1/2)$ transition excited in a TW Hya like protoplanetary disk in an emission surface around $z/r=0.3$, with (\textit{lower row}) and without (\textit{upper row}) a layer of dust present between the emission surfaces. In each subfigure, the spectra for different optical depths (left to right) and azimuthal angles (different colors within figures) are given. The dashed lines are from spectra assuming no magnetic field permeating the disk.}
\label{fig:Stokes_I_50}
\end{figure*}

The broadening of spectral lines may be extracted from the total intensity profiles. We plot the total intensity spectra of the CN $(N,J,F) \to (N',J',F')=(1,1/2,1/2) \to (0,1/2,1/2)$ transition in Fig.~(\ref{fig:Stokes_I_50}), where we plot spectra for varying optical depths, azimuthal angles, and with and without the presence of an optically thick dust layer. Inside the figure, we indicate the simulations assuming negligible magnetic fields, $x_Z \to 0$, by the dashed lines. Comparing those simulations to simulations with magnetic fields, we note a consistent broadening of the line profiles due to the Zeeman effect for the emission emerging from different azimuthal angles. The broadening is most noticeable for optically thin lines, where Zeeman broadened lines clearly show weaker signals, that accordingly have a broader FWHM. From inspection of the spectra, only a weak variation of the broadening with the azimuthal angle is observed. The simulations with the dust layers yield a lower total intensity, but the Zeeman broadening appears to have been affected only to a small degree. We explore this feature in more detail later on.

\begin{figure}[h]
\centering
\includegraphics[width=0.5\textwidth]{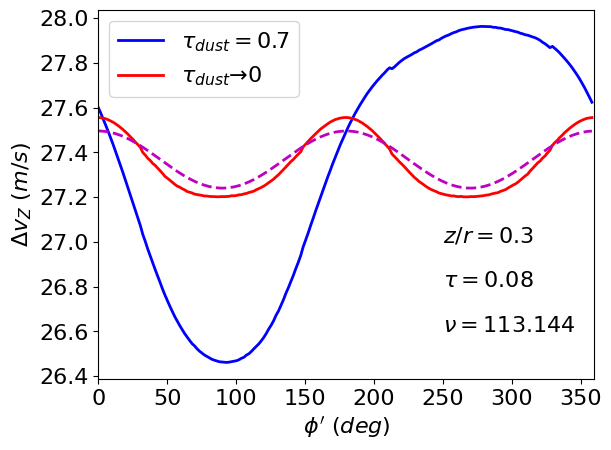}
\caption{Zeeman broadening at deprojected distance $r_c=50$ AU, as a function of the azimuthal angle. Simulations with and without a dust-layer are plotted and the dashed is a fitting function (see text).}
\label{fig:broad_psi}
\end{figure}

We extracted the Zeeman broadening, as defined in Eq.~(\ref{eq:zeeman_broad_parameter}) from the total intensity spectra, and plotted it against the azimuthal angle, for the optically thin simulations, in Fig.~(\ref{fig:broad_psi}). The azimuthal variability of the Zeeman broadening is relatively weak compared to the total broadening: about $5\%$. We note a different variability for the simulations with and without a dusty midplane layer. The difference is due to the $\cos^2 \theta$ term in Eq.~(\ref{eq:proj_simple}) that is dependent on the product between the vertical, $B_{\mathrm{v}}$, and toroidal, $B_{\mathrm{t}}$, magnetic field  and so depends on the polarity of the toroidal magnetic field that changes sign through the midplane. We plot Eq.~(\ref{eq:broad_disk}), putting the polarity-dependent term at zero, in Fig.~(\ref{fig:broad_psi}), where we observe an excellent agreement with the simulations that assume an optically thin dust layer. We acquire a good fit because the absence of a midplane dust layer suppresses the expression of the polarity-dependent term in the Zeeman broadening. The polarity-dependent term does come to expression when a dust layer is present, but its precise effect is difficult to model. However, averaged over all azimuthal angles, as motivated in Section 2.2.4., the polarity-dependence is eliminated, regardless of the presence of a dust layer in the midplane. 

In Table~\ref{tab:rc50_sims}, we list the broadening of the different CN Band $3$ transitions that we explored in our simulations. First, the simulations show that the differently placed emission surfaces at $z/r=0.1$ and $z/r=0.3$ show very different Zeeman broadening for the same transition. This is because of the weaker thermal line broadening at the $z/r=0.1$-surface, as this is situated in cooler gas. Between transitions, we find strong variance of the Zeeman broadening, which for the optically thin simulations of the $z/r=0.3$-surface varies from $96$ m/s to $6.4$ m/s. This may be expected from the quadratic dependence of the broadening on the Zeeman shift, and the variation of the Zeeman splitting factor between the CN transitions. The predicted Zeeman broadening is weakly dependent on the optical depth when lines are $\lesssim 1$ in optical depth. Our estimates of the Zeeman broadening increase for $\tau \gtrsim 1$, but it is difficult to model them with the formalism we introduced in Section 2.

\subsection{Linear polarization}
\begin{figure*}[ht!]
\centering
\includegraphics[width=\textwidth]{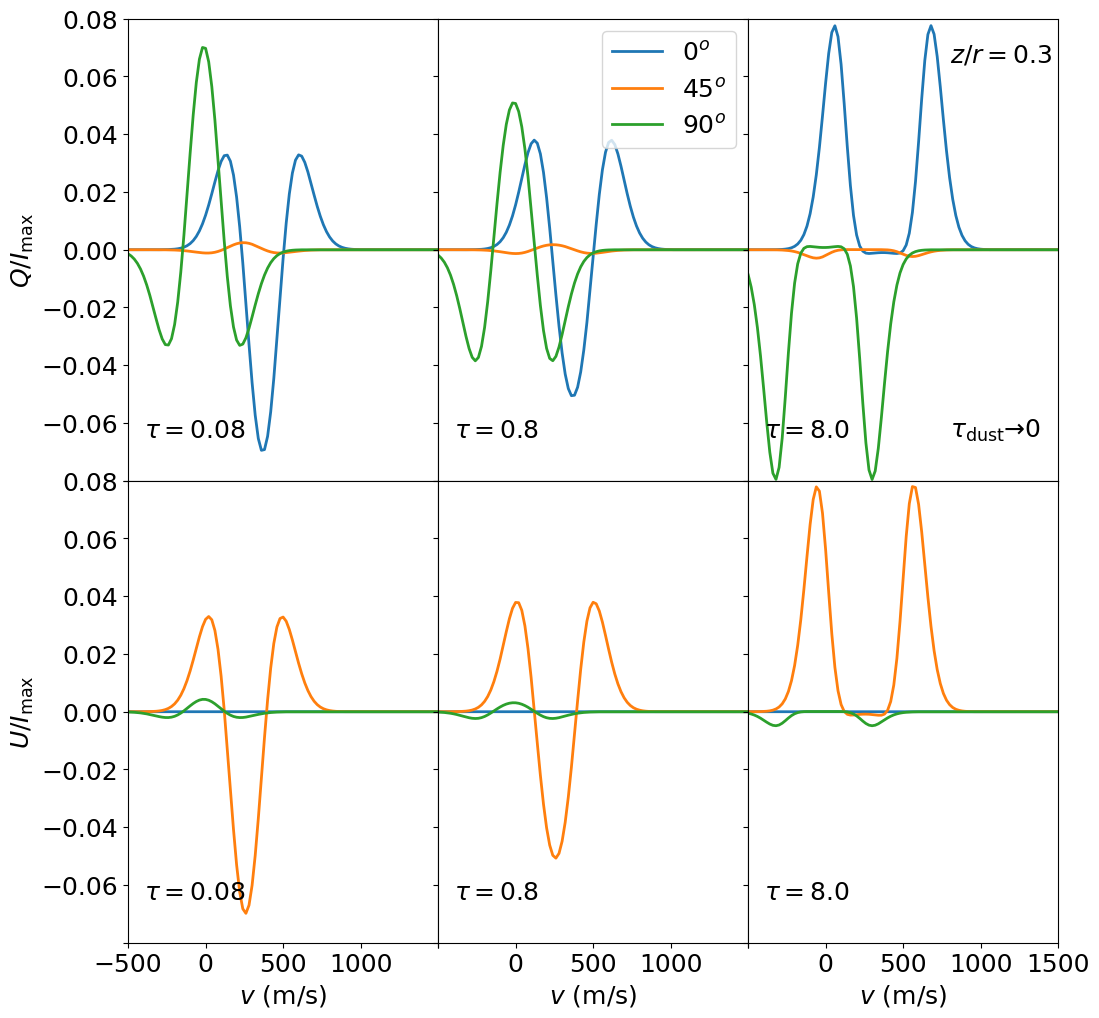}
\caption{Spectra of the linearly polarized intensity of the CN $(N,J,F) \to (N',J',F')=(1,1/2,1/2) \to (0,1/2,1/2)$ transition excited in a TW Hya like protoplanetary disk, in an emission surface around $z/r=0.3$. In the upper row, we plot the Stokes $Q$ spectra and in the lower row we plot the Stokes $U$ spectra.}
\label{fig:Stokes_QU_50}
\end{figure*}
We plot the linear polarization spectra of the CN $(N,J,F) \to (N',J',F')=(1,1/2,1/2) \to (0,1/2,1/2)$ transition in Fig.~(\ref{fig:Stokes_QU_50}). We plot both the Stokes $Q$ and $U$ spectra, assuming an optically thin dust layer, for three azimuthal angles and a varying optical depth. While the Stokes $U$ is almost completely suppressed at $0^o$ and $90^o$, the Stokes $Q$ is suppressed at $45^o$. From our simulations and Eq.~(\ref{eq:qu_fracs}), we note that the linearly polarized Stokes parameters change sign for $\phi' \to \phi'\pm \pi/2$. Also, for simulations $\tau \lesssim 1$, the linear polarization spectra are characterized by a sign change towards the line wings. Therefore, both excellent angular and spectral resolutions are required to resolve the linear polarization signal. Optically thick spectral lines show strong signals toward their line wings, while toward the line center, no polarization is produced. 

\begin{figure}[ht!]
\centering
\includegraphics[width=0.4\textwidth]{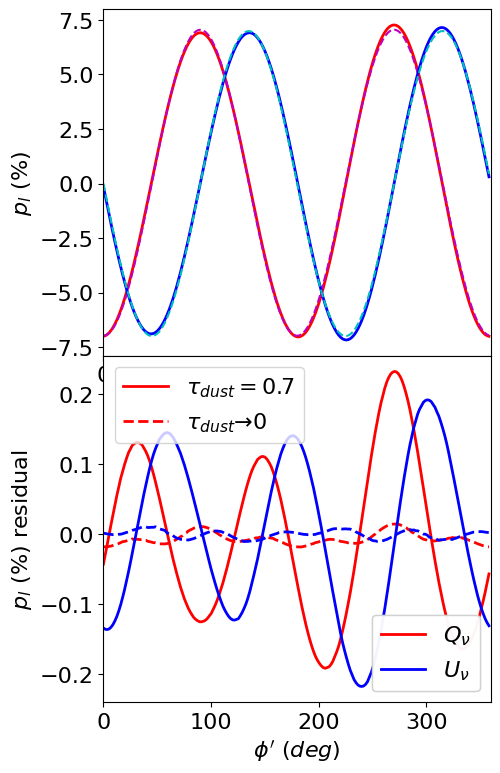}
\caption{(upper) Linear polarization, from the Stokes-$Q$ and -$U$ parameters at the line center, at deprojected distance $r_c=50$ AU and as a function of the azimuthal angle, $\phi'$. Simulations with and without a dust-layer are plotted inside the figure. (lower) residuals from subtracting the fitting function (see text).}
\label{fig:QU_psi}
\end{figure}

In Fig.~(\ref{fig:QU_psi}) we plot the Stokes $Q_{\nu}$ and $U_{\nu}$ linear polarization intensities at the line center, normalized with respect to the total intensity at the line center, $I_{\nu_0}$, as a function of the azimuthal angle from the optically thin simulations. Simulations with and without a dust layer in the midplane are presented, but they differ only slightly. It is readily apparent that the dominant variation with the position angle in the Stokes $Q_{\nu_0}$ adheres to $\propto \cos 2\phi'$, while for the Stokes $U_{\nu_0}$ the variation is as $\propto \sin 2\phi'$. We note that, assuming the polarity-dependent term is zero, Eq.~(\ref{eq:qu_fracs}) describes the linear polarization with high precision. When we subtract Eq.~(\ref{eq:qu_fracs}) from our simulated results we are left with the contributions of the polarity-dependent terms to the linear polarization. We plot the residuals in the lower part of Fig.~(\ref{fig:QU_psi}) where we note that they are a fraction of the total signal and show different profiles between the simulations with and without a dusty midplane layer. Regardless of the presence of the dusty midplane layer, the residuals average to zero, when using the weighting schemes outlined in Section 2.2.4.

We have computed the linear polarization fractions, using Eq.~(\ref{eq:pQ_fac}), of every CN Band $3$ transition, for three optical depths, and listed them in Table~\ref{tab:rc50_sims}. The linear polarization shows a dependence on the optical depth, where the strongest linear polarization is found for optically thin transitions and, above $\tau \sim 1$, progressively diminishes with increasing optical depth. While this is indeed expected for the linear polarization at the line center, we do find that the polarization in the line wings increases with the optical depth. For the optically thin simulations, the expected linear polarization fractions are between $0.0\%$ and $13.0\%$ for the $z/r=0.3$ simulations, and about twice those estimates for the $z/r=0.1$ simulations.

\section{Discussion}
\subsection{Magnetic field detection in protoplanetary disks through spectral line observations}
We have studied the signature of the magnetic field in the Band 3 transitions of CN, excited in a protoplanetary disk. The signature of the magnetic field manifests in the CN spectra through circular and linear polarization, and in the (Zeeman) broadening of the spectral lines. These effects each have a unique relation to the magnetic field of the region that the spectral lines trace. In the following, we discuss how to characterize the underlying magnetic field from observations through each of these effects, and discuss their limitations and uncertainties.


\subsubsection{Circular polarization}
Extracting magnetic field information through circular polarization observations of the Zeeman effect of paramagnetic molecules is widely regarded as one of the most reliable ways to measure the magnetic field strength of astrophysical regions \citep{crutcher:19}. Indeed, the first attempts to characterize protoplanetary disk magnetic fields have used this method, but so far, observations have only been able to set upper limits on the magnetic field properties due to non-detections of the circular polarization \citep{vlemmings:19, harrison:21}.

The production of circular polarization in spectral lines scales with the Zeeman shift in Doppler units, $x_Z$, while the effects of Zeeman broadening and linear polarization scale with $x_Z^2$. Considering that for protoplanetary disks $x_Z<1$, one would expect the circular polarization to be the most sensitive magnetic field tracer. However, the dominant toroidal (and radial) components of the magnetic field likely change sign across the disk midplane, thus suppressing the circular polarization produced by these magnetic field components \citep[this was also pointed out in ][]{mazzei:20}. Additionally, for weakly inclined disks, primarily the vertical component of the magnetic field comes to expression in the circular polarization, which is expected to be a factor $\sim 10$ weaker than the dominant toroidal magnetic field component \citep{bethune:17}.

We considered two mechanisms through which the cancellation of circular polarization due to the midplane polarity change can be diminished. First, when a velocity shift between both sides of the disk is present, the circular polarization profiles of the two disk-sides are non-resonant and circular polarization is recovered in proportion to the size of the velocity shift. Second, with the presence of an optically thick dust layer in the midplane, a part of the circular polarization of the back side of the disk is absorbed and the cancellation of circular polarization is accordingly partially negated.

We apply our results to the existing circular polarization measurements of CN towards protoplanetary disks. \citet{vlemmings:19} observed a subset of the Band $6$ transitions of CN towards TW Hya using ALMA. They obtain the tightest constraints on the vertical magnetic field through performing of a stacking of the CN transitions in combination with an azimuthal stacking. The toroidal magnetic field component they extract through subtracting the stacked results between the red- and blue-shifted parts of the major axis. We performed 3D polarized radiative transfer simulations of a TW Hya like disk and found that the polarization fraction was reproduced excellently by Eq.~(\ref{eq:pV_dust}). From this result, we have no comments on the method to estimate the vertical component of the magnetic field, but point out that the procedure meant to extract the toroidal magnetic field yields $\tanh \frac{\tau_{\mathrm{dust}}}{2}B_{\mathrm{t}}$, rather than $B_{\mathrm{t}}$. With current estimates of the dust opacity of TW Hya \citep{macias:21}, we may put $\tanh \frac{\tau_{\mathrm{dust}}}{2} \approx 0.46$. On the basis of our comments on the extraction method and our radiative transfer simulations, we suggest updating the constraint of \citet{vlemmings:19} of the toroidal magnetic field to $B_{\phi} < 65 \ \mathrm{mG}$. Additionally, we point out that the procedure to extract the toroidal magnetic field strength using the stacking of spectral lines along the line of nodes may be improved by a positional angle stacking with the appropriate weighing of $\cos \phi'$ (see Eq.~(\ref{eq:pV2}). This analysis will be undertaken in a future paper. 

\citet{harrison:21} observed a subset of CN Band $6$ transitions towards AS 209 in search of circular polarization. Using a similar method as \citet{vlemmings:19} -- stacking the transitions of both the red- and blue-shifted parts of the disk -- they extract a toroidal magnetic field strength of $<8.7$ mG, but note that this may be an underestimation due to cancellation effects. First, the method that \citet{harrison:21} used to extract the toroidal magnetic field is by stacking the circularly polarized emission in either the red- and blue-shifted parts of the disk. According to Eq.~(\ref{eq:pV2}), this procedure extracts $2/\pi B_{\mathrm{t}}$ times the toroidal magnetic field, while \citet{harrison:21} assumed it extracted $B_{\mathrm{t}}$. Therefore, their estimated toroidal magnetic field strength has to be raised with a factor $\pi/2$. Additionally, the estimates of the toroidal magnetic field are impacted by cancellation effects. AS 209 is inclined at $35.3^o$ \citep{fedele:18}, and between the disk major- and minor-axes, a velocity-shift of the order of the Doppler width is expected \citep{teague:18b}. For such velocity-shift, only weak circular polarization cancellation effects are expected, but note that around the major-axes, where the toroidal projection is maximal, no velocity shift occurs. Towards the major-axes, based on estimates of the dust optical depth \citep{perez:12, tazzari:16}, we can put the relevant dust optical depth at $\tau_{\mathrm{dust}} \sim 0.2$, and find that dust absorption recovers $\sim 10\ \%$ of the circular polarization due to the toroidal magnetic field there. Based on these considerations, we conservatively estimate that about $40 \%$ of the circular polarization due to the toroidal magnetic field is recovered, and accordingly, we suggest updating their constraint of the toroidal magnetic field strength to $<35$ mG. 

We have established that circular polarization observations can be reliably employed to extract information on the vertical magnetic fields of protoplanetary disks. However, the emergence of circular polarization due to the toroidal magnetic field is significantly affected by the sign change of the magnetic field between both sides of the disk \citep[see also, ][]{mazzei:20}. Accordingly, the circular polarization due to the toroidal magnetic field is partially cancelled, and the interpretation of circular polarization signals for their magnetic field information is contingent on accurate knowledge of the dust optical depth in the disk, as well as the velocity-structure and emission surface. Unavoidably, this places large uncertainties on estimates of the toroidal magnetic field using circular polarization observations.

\subsubsection{Zeeman broadening}
The splitting of spectral lines through the Zeeman effect also broadens them. We established that the Zeeman broadening is proportional to the square of the total magnetic field strength, in addition to the square of the line-of-sight component of the magnetic field. Each of these contributions is proportional to the factors $\bar{Q}$ and $\Delta Q$, which we have listed for the CN transitions in Table \ref{tab:CN_mol}. We showed that Eq.~(\ref{eq:zeeman_broad}) accurately describes the Zeeman broadening of weakly Zeeman split, optically thin emission lines, while for very optically thick lines or strong magnetic fields ($x_Z>0.3$), Eq.~(\ref{eq:zeeman_broad}) tends to underestimate the broadening. In the case of TW Hya, \citet{teague:20} estimated CN Band $3$ transitions to have $\tau \lesssim 1$, while current magnetic field limits place the Zeeman splitting well under $x_Z<0.3$, so we may establish that Eq.~(\ref{eq:zeeman_broad}) describes the Zeeman broadening of CN lines excited in this disk with high accuracy. 

We showed that the position dependent Zeeman broadening is well described by Eq.~(\ref{eq:broad_disk}) if one disregards the contribution to the Zeeman broadening due to the changing polarity of the toroidal magnetic field through the midplane. The contribution of the polarity dependent cross-term to the Zeeman broadening suggests that its interpretation, just as circular polarization observations, is contingent on accurate modeling of the disk radiative transfer. Strictly, if aiming for a resolved profile of the magnetic field of a strongly inclined disk, this is the case. However, taking into consideration the expected axisymmetry of the disk magnetic field, one may devise averaging schemes that extract the axisymmetric magnetic field components, $B_{\mathrm{v}}$ and $B_{\mathrm{t}}$, while eliminating any influence of the cross-terms on the averaged quantities. Our radiative transfer simulations confirmed Eq.~(\ref{eq:broad_average}), and showed that when one stacks the Zeeman broadening over the range of position angles, then the resulting averaged quantity is insensitive to the polarity-dependent terms in $\cos^2 \theta$ and thus insensitive to any cancellation effects of the Zeeman broadening between both sides of the disk. Thus, extracting magnetic field information from the Zeeman broadening in this way, does not require knowledge of the continuum absorption or a velocity shift between the two emission surfaces along the line-of-sight. Moreover, the result of Eq.~(\ref{eq:broad_average}) retains its quality for disks with varying inclination.

In order to separate the Zeeman broadening from other broadening mechanisms, one needs to observe a number of lines with varying Zeeman parameters simultaneously. A molecule such as CN, which exhibits several manifolds of transitions that lie relatively close in frequency, is ideally suited for this. Our simulations indicate, that at a deprojected distance of $r_c = 50$ AU, assuming an emission surface at $z/r=0.3$ and a magnetic field such as Eq.~(\ref{eq:B_field}), that the $9$ transitions of the $N=1-0$ manifold vary in Zeeman broadening, from $6.4$ m/s to $96$ m/s. 

Detection of such a level of broadening should be possible with current facilities. This can be seen by considering the accuracy of Gaussian fits to noisy spectra described in Appendix B of \citet{teague:22}. The SNRs reported for CN $N = 1-0$ emission from TW Hya in \citet{teague:20} vary from 2 to 50, depending on the specific hyperfine component. The sampling rate, defined as the ratio of the FWHM to the channel spacing, is $\approx 3$, assuming a FWHM of $250~{\rm m\,s^{-1}}$ and adopting the highest channel sampling of 30~kHz, or $80~{\rm m\,s^{-1}}$ at the frequency of the $N=1-0$ transition. From Figure~13 from \cite{teague:22}, this yields an accuracy on the line width of $50~{\rm m\,s^{-1}}$ for the low SNR lines, and down to $13~{\rm m\,s^{-1}}$ for the higher SNR lines. We note that while for an individual component this precision is insufficient to detect Zeeman broadening, measuring the line widths of the 9 hyperfine components simultaneously will yield a ${\sim} \sqrt{9}$ improvement in the accuracy. This should be sufficient to detect the predicted broadening of between $6.4~{\rm m\,s^{-1}}$ and $96~{\rm m\,s^{-1}}$. Longer integrations than the 78~minute observations reported in \citet{teague:20} would allow for smaller changes in line width, associated with weaker magnetic fields, to be detected. Looking to the future, the ALMA2030 Wideband Sensitivity Upgrade will achieve a velocity resolution of at least $10~{\rm m\,s^{-1}}$ at 113~GHz \citep{carpenter:23}, providing another ${\sim}3$ improvement in the accuracy of line width measurements and opening up the study of Zeeman broadening to a wide variety of sources.
ee


\subsubsection{Linear polarization}
Zeeman split lines produce linear polarization in proportion to the square of the plane-of-the-sky component of the magnetic field. To predict the total linear polarization fraction for the frequently observed CN transitions from protoplanetary disks, we derived Eqs.~(\ref{eq:pQ_fac}) and (\ref{eq:qu_fracs}) which we showed compare well to full polarized radiative transfer simulations with optically thin emission. As the linear polarization is dependent on the plane-of-the-sky magnetic field, we predict that weakly inclined disks produce the strongest linear polarization, because the dominant toroidal magnetic field comes best to expression here. We compute that the CN Band $3$ transitions, assuming a TW Hya like disk at deprojected distance of 50~AU and a magnetic field as Eq.~(\ref{eq:B_field}), exhibit polarization levels up to $10 \ \%$---far exceeding the modeled $\sim 1\%$ polarization levels due to circular polarization.  

Linear polarization observations commonly return the two Stokes parameters, $Q_{\nu}$ and $U_{\nu}$, which span a two-dimensional quasi vector on the plane of the sky and may be related to the magnetic field projected onto the plane-of-the-sky. In Eqs.~(\ref{eq:qu_fracs}), we related the linear polarization fractions of both Stokes parameters to the magnetic field properties of the disk. We noted that both Stokes parameters have a dependence on the square of the vertical and toroidal components of the magnetic field, but also on their cross elements, $B_{\mathrm{v}}B_{\mathrm{t}}$, which change sign between both back and front side of the disk. Our radiative transfer simulations indicate that the impact of the cross-terms on the linear polarization is difficult to estimate when an absorbing dust layer is present and when the disk is observed at high inclination. For such sources, the interpretation of a resolved linear polarization mapping requires accurate modeling of the disk radiative transfer. 

However, one can exploit the axisymmetry of the disk to extract different magnetic field components from the linear polarization data by applying the weighted stacking procedure as outlined in Section 2.2.4.,~to obtain the 3D magnetic field components. Our modeling shows that the stacked quantities are insensitive to the polarity-dependent terms that require detailed modeling to be quantified. Thus, they provide for a robust procedure to extract magnetic field information from observations. We note that in order to perform the suggested azimuthal stacking, excellent spatial and spectral resolution is requried. Spatial resolution is an important factor as background gradients in the gas temperature and velocities can result in `beam smearing', as discussed in \citet{teague:16}. A good rule of thumb is that the angular resolution must be such that all properties are expected to be broadly constant across the beam size. Spectral resolution is important as it allows one to perform the stacking while correcting for the disk kinematics. It is for this reason that interferometers such as ALMA, that provide a combination of excellent angular and spectral resolution, are key to these observations. 

The (sub-)millimeter continuum emission emerging from protoplanetary disks is linearly polarized up to a few percents \citep{stephens:17, vlemmings:19}. The precise polarization mechanism is under active debate, and is dependent on the wavelength, dust and disk properties, and on the dust alignment mechanism \citep{kataoka:15, kataoka:17, stephens:17, tazaki:17, hoang:22}. We estimate the impact of the continuum polarization on the emergent spectral line polarization by considering the polarization resolved emissivity and extinction properties of the midplane dust. First, we estimate the impact of the continuum emission on the spectral line polarization signal. The continuum emission is about $10$ times weaker than co-spatial CN emission toward the line center. In addition, the continuum emission that adds to the spectral line signal will be weakened by a factor $e^{-\tau_{\mathrm{CN}}} \sim 1/2$, due to absorption in the front side emission surface. Putting the continuum polarization fraction at 5\%, we thus estimate its contribution to the spectral line linear polarization fraction to be $0.25 \%$. The spectral line polarization can also be impacted through extinction in the midplane. The extinction properties of midplane dust include a conversion of Stokes I to Q, whose efficiency depends on the degree of alignment of the midplane dust \citep{andersson:15}. If we take the conversion of Stokes I to Q to be of the order of a percent of the total extinction, we may estimate the contribution of the converted back side CN emission to linearly polarized radiation as $1\% / (e^{-\tau_{\mathrm{dust}}}+1) \sim 0.1-0.5 \%$. However, since this polarized radiation will be subsequently absorbed in the front side emission surface, it will get suppressed by an additional factor $e^{-\tau_{\mathrm{CN}}} \sim 1/2$, to yield $\sim 0.05-0.25 \%$. In summary, we estimate an error of $\sim 0.35\%$ in the spectral line linear polarization fractions due to the polarized emission and absorption properties of the protoplanetary disk continuum.

The observations of \citet{vlemmings:19} and \citet{harrison:21} were performed in ALMA full polarization mode, and also yielded linear polarization data. Both observations were performed in Band $6$, where linear polarization levels are expected to be lower than for CN Band $3$ transitions. Still, the linear polarization of the CN spectral lines may be advantageously used to constrain the magnetic field strength using the aforementioned methods. We leave such an analysis for a future publication.

\subsection{Zeeman effect in spectral lines}
The measurement of the Zeeman effect is widely recognized as one of the most reliable tracers of the magnetic field strength in astrophysics \citep{crutcher:19, semel:89}. Circular polarization measurements to interstellar medium (ISM) molecules have allowed for the measurement of the (line-of-sight component of the) magnetic field in star-forming regions \citep{crutcher:19}, at a range of densities and scales from the diffuse ISM \citep{heiles:05} to masers excited close to (massive) protostars \citep{vlemmings:06a, vlemmings:08, lankhaar:18}. In this paper, we have laid particular emphasis on the linear polarization and line broadening that is associated with the Zeeman effect. 

The linear polarization and broadening properties of Zeeman split spectral lines are often overlooked, as the Zeeman splitting of ISM spectral lines is generally weak compared to their line width (circular polarization fractions of $p_V \sim 0.01$ are observed). Based on the discussion of \citet{crutcher:93} about the radiative transfer of polarized radiation due to Zeeman effects, one expects the linear polarization and Zeeman broadening to be on the order of $p_V^2$, which would yield no detectable signature. We point out, though, that \citet{crutcher:93} assume no intragroup spread of Zeeman shifts within the groups of $\sigma^{\pm}$- and $\pi^0$-transitions. This is a good approximation for transitions with low angular momentum (and perfect for $F=1-0$ transitions), but becomes increasingly unreliable with increasing angular momentum of the quantum states associated with the transition. We showed in Section~\ref{sec:theory_zeeman} that the intragroup spread in frequency of the individual magnetic sublevel transitions impacts the estimates for the linear polarization and Zeeman broadening significantly. This impact is conveniently represented by the factors $\Delta Q$ for the linear polarization and additionally the factor $\bar{Q}$ for the Zeeman broadening. In Fig.~(\ref{fig:zeeman_band3}) we plotted transitions that exhibit a low and high intragroup spread in Zeeman shifts. For transitions with high angular momentum, the $Q$-factors can assume values $>1000$ (see Table \ref{tab:CN_mol}). Accounting for the intragroup spread in Zeeman shifts significantly impacts the expected linear polarization and Zeeman broadening compared to estimates assuming no intragroup spread of frequencies ($\bar{Q}=-\Delta Q=1$). 

Transitions that exhibit strong Zeeman broadening or linear polarization may be more sensitive tracers of the magnetic field strength than circular polarization. Indeed, there are distinct advantages to using Zeeman broadening as a magnetic field strength tracer. Firstly, full polarization observations are technically challenging; they require excellent weather conditions to be calibrated, often at the costs of other observational parameters such as the spectral resolution. Secondly, the Zeeman broadening and linear polarization are a direct probe of the magnetic energy of the probed region. Thirdly, due to their quadratic dependence on the magnetic field, Zeeman broadening or linear polarization will not suffer from a loss of signal, as would circular polarization, under a variable magnetic direction along the line-of-sight; a favorable property when tracing magnetic fields in protoplanetary disks. In turbulent regions, such as molecular clouds, this is highly advantageous over circular polarization measurements. The observation of several Zeeman induced effects is highly complementary, with the simultaneous observation of circular polarization and linear polarization yielding the 3D direction and strength of the magnetic field of the probed region. 



\section{Conclusions}
The most reliable and direct method of magnetic field detection is through detection of the Zeeman effect in spectral lines. We have considered the detection of the Zeeman effect in the spectral lines of CN that are excited in protoplanetary disks using (polarized) radiative transfer modeling. While previous attempts to detect and characterize protoplanetary disk magnetic fields have been via circular polarization observations \citep{vlemmings:19, harrison:21}, we dedicated particular attention to the magnetic field signature in Zeeman broadening and in the linear polarization of Zeeman split lines. In order to characterize the observability and robustness regarding magnetic field characterization of these observable features of the Zeeman effect, we performed both simplified and detailed modeling of the transfer of polarized radiation of Zeeman split CN spectral lines that are excited in protoplanetary disks. 

We find that the magnetic field sign inversion through the disk midplane, characteristic of protoplanetary disks, significantly impacts the circular polarization signal. Because of the magnetic field inversion along the line-of-sight, much of the circular polarization that is due to the toroidal magnetic field is lost. The reduction in circular polarization can be partially suppressed through a velocity shift between the back and front side emission surfaces, or an optically thick dust layer. Besides negatively impacting the observability, the resulting interpretation of circularly polarized signals for their magnetic field information is contingent on accurate modeling of these effects. Application of our results to the existing circular polarization observations of TW Hya \citep{vlemmings:19} and AS 209 \citep{harrison:21} suggests raising their upper limits on the toroidal magnetic field strength from $30$ mG to $65$ mG, and from $8.7$ mG to $35$ mG, respectively.

The production of linear polarization and Zeeman broadening scale quadratically with the magnetic field strength, while circular polarization scales linearly with the line-of-sight component of the magnetic field. From our radiative transfer simulations we found that for moderately inclined disks the observation of linear polarization and Zeeman broadening carry two advantages over circular polarization measurements. First, due to the quadratic dependence on the magnetic field strength, linear polarization and Zeeman broadening are only weakly affected by the magnetic field inversion along the line-of-sight. Second, linear polarization and Zeeman broadening are sensitive to the plane-of-the-sky magnetic field and the total magnetic field strength, respectively, meaning that they are sensitive to the toroidal magnetic field, which is expected to be the dominant component of the magnetic field.

We have presented a method to interpret and predict linear polarization and Zeeman broadening observations. Importantly, we introduced the coefficients $\bar{Q}$ and $\Delta Q$, that take into account the enhanced Zeeman broadening and linear polarization due to intragroup spread of Zeeman shifted transitions, and which are omitted in the seminal work of \citet{crutcher:93}. With the $\bar{Q}$ and $\Delta Q$ coefficients, simplified radiative transfer models could reproduce synthetic observations from full 3D polarized radiative transfer modeling with high fidelity. Conversely, the simplified radiative transfer models may thus be used in the interpretation of observations. From the simultaneous observation of the linear polarization, and the Zeeman broadening or circular polarization, the magnetic field strength and its 3D direction may be derived. We predict that such observations are feasible for a moderately inclined protoplanetary disk such as TW Hya.

\begin{acknowledgements}
BL acknowledges support for this work from the Swedish Research Council (VR) under grant number 2021-00339. Simulations were performed on resources at the Chalmers Centre for Computational Science and Engineering (C3SE) provided by the Swedish National Infrastructure for Computing (SNIC). 
\end{acknowledgements}

\bibliography{lib.bib}

\begin{appendix}
\section{Polarized radiative transfer of Zeeman splitted lines}
\label{sec:ap_A}
We consider the transfer of (polarized) radiation in a Zeeman split spectral line. We relate the polarized propagation properties of the Zeeman split spectral line to its propagation properties in the limit of no Zeeman splitting, where we let the absorption coefficient and source function be $\kappa_{\nu}$ and $S_{\nu_0}$. We factorize the absorption coefficient $\kappa_{\nu} = k_0 \bar{\phi}(x)$, in an absorption constant and a dimensionless line-profile, where we use Doppler-normalized units for the line-profile: $x=\frac{\nu-\nu_0}{\Delta \nu_D}$, with $\nu$ as the frequency, $\nu_0$ as the frequency of the line center and $\Delta \nu_D$ as the Doppler broadening in frequency units. Even though we defined the line profile as a Gaussian in the main text, our following discussion is appropriate for any profile function, provided that $\int dx \ \bar{\phi}(x) = 1$.

When a magnetic field is present, the spectral line under consideration splits up into a multitude of transitions between the individual magnetic sublevels of the upper and lower state. As in the main text, we consider a transition between two states, with angular momenta $F_1$ and $F_2$ for the upper and lower states respectively. Transitions between magnetic sublevels are shifted in frequency according to Eq.~(\ref{eq:zeeman_shift}). Accordingly, the line profile of the transition $\ket{F_1 m_1} \to \ket{F_2 m_2}$ is Zeeman shifted, $\bar{\phi} (x + x_B [g_1 m_1 - g_2 m_2])=\bar{\phi}_{m_1,m_2}$, where $x_B = \mu_B B / h \Delta \nu_D$ is the Zeeman shift term in Doppler units.

As discussed in the text, transitions between magnetic sublevels can be divided into three groups, the $\pi^0$- and the $\sigma^{\pm}$-transitions, that are associated with the $\Delta m=0$ and $\Delta m=\pm 1$ transitions. The individual transition groups emit polarized emission and have opacities that are dependent on the angle between the propagation direction and the magnetic field that gives rise to the Zeeman splitting (see also main text). We therefore define a total line profile per transition group,
\begin{align}
\label{eq:line}
\bar{\phi}_q(x,x_B) = \sum_{m_1} S_q(F_1,F_2,m_1) \bar{\phi}_{m_1,m_1+q},
\end{align}
where $S_q(F_1,F_2,m_1)$ is the relative line-strength of the $\ket{F_1 m_1} \to \ket{F_2 m_1+q}$ transition \citep[see also main text and equation 3.16 of][]{landi:06}. We now cite the polarized radiative transfer equation, over an incremental distance, $ds$, of a Zeeman split spectral line \citep{landi:06,rees:89},
\begin{align}
\label{eq:full_pol}
\frac{d}{ds} \begin{pmatrix}I_{\nu} \\ Q_{\nu} \\ U_{\nu} \\ V_{\nu} \end{pmatrix} =
- \begin{pmatrix}
\kappa_{I} & \kappa_{Q} & \kappa_{U} & \kappa_{V} \\
\kappa_{Q} & \kappa_{I} & \kappa_{V}'&-\kappa_{U}'\\
\kappa_{U} &-\kappa_{V}'& \kappa_{I} & \kappa_{Q}'\\
\kappa_{V} & \kappa_{U}'&-\kappa_{Q}'& \kappa_{I}
\end{pmatrix}
\begin{pmatrix}I_{\nu}-S_{\nu_0} \\ Q_{\nu} \\ U_{\nu} \\ V_{\nu} \end{pmatrix},
\end{align}
where the Stokes parameters are the same as those in the main text. The polarized absorption coefficients are,
\begin{subequations}
\label{eq:pol_abs}
\begin{align}
\kappa_I &= \frac{k_0}{2} \left[\bar{\phi}_0 \sin^2 \theta + \frac{\bar{\phi}_{1} + \bar{\phi}_{-1}}{2} \left(1+\cos^2 \theta \right) \right], \\
\kappa_Q &= \frac{k_0}{2} \left[\bar{\phi}_0 -  \frac{\bar{\phi}_{1} + \bar{\phi}_{-1}}{2} \right] \sin^2 \theta \cos 2\eta , \\
\kappa_U &= \frac{k_0}{2} \left[\bar{\phi}_0 -  \frac{\bar{\phi}_{1} + \bar{\phi}_{-1}}{2} \right] \sin^2 \theta \sin 2\eta , \\
\kappa_V &= \frac{k_0}{2} \left[\bar{\phi}_{1} - \bar{\phi}_{-1}\right] \cos \theta ,
\end{align}
\end{subequations}
and the $\kappa_{Q,U,V}'$ terms are transformation coefficients that are related to the $\kappa_{Q,U,V}$ opacities through the Kramers-Kronig relations.

We now endeavour to obtain simplified polarized absorption coefficients. To this end, we represent the line profiles of the transition groups by a Taylor expansion,
\begin{align}
\label{eq:line_taylor}
\bar{\phi}_q = \sum_n \left[\frac{d^n \bar{\phi}_q}{d x_B}\right]_{x_B=0} \frac{x_B^n}{n!},
\end{align}
around the Zeeman shift term, $x_B$, where we recognize that this representation is best applied in the weak-field limit, to lines with $x_B < 1$ \citep{landi:06}. We combine Eqs.~(\ref{eq:line}) and (\ref{eq:line_taylor}), and note
\begin{align}
\label{eq:taylor_dev}
\left[\frac{d^n \bar{\phi}_q}{d x_B^n}\right]_{x_B=0} = \sum_{m_1} S_q(F_1,F_2,m_1) \left[\frac{d^n \bar{\phi}_{m_1,m_1+q}}{d x_B^n}\right]_{x_B=0}.
\end{align}
From the definition of $\bar{\phi}_{m_1,m_1+q}$, we recognize that we may substitute the differential $dx_B = dx' / [g_1 m_1 - g_2 (m_1+q)]$, where $x'=x+x_B [g_1 m_1 - g_2 (m_1+q)]$. Furthermore, $x' \to x$, in the limit of $x_B \to 0$, so 
\begin{align}
\label{eq:taylor_dev_sub}
\left[\frac{d^n \bar{\phi}_{m_1,m_1+q}}{d x_B^n}\right]_{x_B=0} = (g_1 m_1 - g_2 [m_1+q])^n \frac{d^n \bar{\phi} (x)}{d x^n},
\end{align}
where $\frac{d^n \bar{\phi}(x)}{d x^n}$ is the $n$-th derivative of the unsplit line profile. We implement Eq.~(\ref{eq:taylor_dev_sub}) in Eq.~(\ref{eq:taylor_dev}), to find,
\begin{align}
x_B^n \left[\frac{d^n \bar{\phi}_q}{d x_B^n}\right]_{x_B=0} &= x_B^n\sum_{m_1} S_q(F_1,F_2,m_1) \nonumber \\ & \times (g_1 m_1 - g_2 [m_1+q])^n\frac{d^n \bar{\phi} }{d x^n}\nonumber \\
&= G_q^{(n)} x_B^n \frac{d^n \bar{\phi} }{d x^n},
\end{align}
and define g-factors, $G_q^{(n)}$, that encapsulate the effects of the Zeeman shifts on the line profile. Truncating the Taylor expansion at $n=2$, while using that $G_0^{(1)}=0$, $G_{1}^{(1)} = -G_{-1}^{(1)}$ and, $G_{1}^{(2)} = G_{-1}^{(2)}$ \citep[these relations follow from the symmetry relations of the line-strength factors, see also][]{landi:06}, we have for the transition group line profiles,
\begin{subequations}
\begin{align}
\bar{\phi}_0 &\simeq \bar{\phi} + \frac{x_B^2}{2} G_0^{(2)} \bar{\phi}'', \\
\bar{\phi}_{\pm 1} &\simeq \bar{\phi} \pm x_B G_{1}^{(1)} \bar{\phi}' + \frac{x_B^2}{2} G_{1}^{(2)} \bar{\phi}'',
\end{align}
\end{subequations}
where we used the short-hand notation for the first- and second-derivatives. Using the approximate expressions for the line profiles in Eq.~(\ref{eq:pol_abs}), we retrieve,
\begin{subequations}
\label{eq:pol_abs_simple}
\begin{align}
\kappa_I/k_0 &\simeq \bar{\phi} + x_B^2 \left[ \frac{G_0^{(2)}+G_{1}^{(2)}}{4} - \frac{G_0^{(2)}-G_{1}^{(2)}}{4} \cos^2 \theta  \right]\ \bar{\phi}'' \\
\kappa_Q/k_0 &\simeq x_B^2 \frac{G_0^{(2)}-G_{1}^{(2)}}{4} \sin^2 \theta \cos 2\eta \ \bar{\phi}'' \\
\kappa_U/k_0 &\simeq x_B^2 \frac{G_0^{(2)}-G_{1}^{(2)}}{4} \sin^2 \theta \sin 2\eta \ \bar{\phi}'' \\
\kappa_V/k_0 &\simeq x_B G_0^{(1)} \cos \theta \ \bar{\phi}'.
\end{align}
\end{subequations}
To correlate our discussion to the literature of radio-astronomical Zeeman observations, we related the proportionality constants to the relative production of circular polarization, through the parameter, $x_Z = x_B G_0^{(1)}$. Thus, we defined $\bar{Q} = (G_0^{(2)}+G_{1}^{(2)}) / (G_0^{(1)})^2$ and $\Delta Q = (G_0^{(2)}-G_{1}^{(2)}) / (G_0^{(1)})^2$ that are the second-order g-factors normalized to the first-order g-factor. In Eq.~(\ref{eq:zeeman_fo}), we give the definition of $x_Z$, where we have used the expression for $G_1^{(1)}$, while in Eqs.~(\ref{eq:zeeman_so}) we give isomorphic definitions of $Q^0=G_0^{(2)}/ (G_0^{(1)})^2$ and $Q^{\pm} = G_{1}^{(2)}/ (G_0^{(1)})^2$.

Using the approximate propagation coefficients of Eq.~(\ref{eq:pol_abs_simple}) in conjunction with the radiative transfer equation of Eq.~(\ref{eq:full_pol}) in the optically thin limit, we retrieve Eqs.~(\ref{eq:zeeman_thin}).

\section{Detailed derivations of Eq.~(15) and (17)}
\label{sec:ap_B}
To derive Eq.~(\ref{eq:pV_vel}), we divide the radiative transfer through the disk into two optically thin propagations through the back- and front sides of the disk emission surfaces. The emission surfaces have equal source function and optical depth $S_{\nu_0}$ and $\tau_{\nu_0}$. We consider the case where $x_Z^2 \ll 1$, and we consider a Doppler normalized velocity shift between the emission surfaces of $\Delta x$, where also, $(\Delta x)^2 \ll 1$. The emergent total intensity is then,
\begin{align}
I_{\nu} &= S_{\nu_0} \tau_{\nu_0} [\bar{\phi}(x-\frac{\Delta x}{2}) + \bar{\phi}(x+\frac{\Delta x}{2})], \nonumber \\ 
        &= 2S_{\nu_0} \tau_{\nu_0} \bar{\phi}(x) + S_{\nu_0} \tau_{\nu_0} \nonumber \\ &\times \left[-2 \bar{\phi}(x) + \bar{\phi}(x-\frac{\Delta x}{2}) + \bar{\phi}(x+\frac{\Delta x}{2}) \right], \nonumber \\
        &\simeq 2S_{\nu_0} \tau_{\nu_0} \bar{\phi}(x) + S_{\nu_0} \tau_{\nu_0} \frac{(\Delta x)^2}{4} \bar{\phi}''(x), \nonumber \\
\label{eq:totI_15}
        &\simeq 2S_{\nu_0} \tau_{\nu_0} \bar{\phi}(x),
\end{align}
while the emergent circular polarization, due to the toroidal magnetic field, with a projection $\pm \hat{\boldsymbol{\phi}}\cdot \hat{\boldsymbol{n}}_{\mathrm{los}}$ in the front and back side of the disk respectively, is,
\begin{align}
V_{\nu} &= -S_{\nu_0} \tau_{\nu_0} x_Z (\hat{\boldsymbol{\phi}}\cdot \hat{\boldsymbol{n}}_{\mathrm{los}}) [\bar{\phi}'(x-\frac{\Delta x}{2}) - \bar{\phi}'(x+\frac{\Delta x}{2})], \nonumber \\ 
\label{eq:totV_15}
        &\simeq S_{\nu_0} x_Z (\hat{\boldsymbol{\phi}}\cdot \hat{\boldsymbol{n}}_{\mathrm{los}}) \Delta x \frac{d \bar{\phi}'(x)}{dx}. 
\end{align}
By using the total intensity from Eq.~(\ref{eq:totI_15}) to substitute the line-profile, $\bar{\phi}$ in Eq.~(\ref{eq:totV_15}), we retrieve Eq.~(\ref{eq:pV_vel}).

To derive Eq.~(\ref{eq:pV_dust}), we divide the radiative transfer through the disk into three propagations. First, we have an optically thin propagation through the back side of the disk. This is followed by a propagation through a dusty midplane. Finally, we have a propagation through the optically thin front side of the disk. The emission surfaces have equal source function and optical depth $S_{\nu_0}$ and $\tau_{\nu_0}$ and the dusty midplane has an optical depth of $\tau_{\mathrm{dust}}$ and a source function $S_{\mathrm{dust}}$. It will later turn out that the  our final result is independent on the dust source function. We consider the case where $x_Z^2 \ll 1$, and we consider a Doppler normalized velocity shift between the emission surfaces of $\Delta x$, where also, $(\Delta x)^2 \ll 1$. The emergent total intensity is then,
\begin{align}
I_{\nu} \simeq S_{\nu_0} \tau_{\nu_0} \bar{\phi}(x) (1+e^{-\tau_{\mathrm{dust}}}) + S_{\mathrm{dust}} (1-e^{-\tau_{\mathrm{dust}}}),
\end{align}
where we have assumed the broadening due to the velocity shift zero, which we motivated in more detail in our derivation of Eq.~(\ref{eq:pV_vel}). Emission from the back side is partially absorbed by the dusty midplane layer, while the emission from the dusty midplane has no significant absorption from the front side of the disk, as we have assumed the emission surfaces optically thin. In fact, the emission from the dusty midplane is removed from the emergent line intensity, as we are interested in the continuum subtracted line intensity
\begin{align}
I_{\nu}^{\mathrm{c.s.}} \simeq S_{\nu_0} \tau_{\nu_0} \bar{\phi}(x) (1+e^{-\tau_{\mathrm{dust}}}).
\end{align}
Taking into account both the midplane absorption and the velocity shift, the emergent circular polarization is
\begin{align}
V_{\nu} &= -S_{\nu_0} \tau_{\nu_0} x_Z (\hat{\boldsymbol{\phi}}\cdot \hat{\boldsymbol{n}}_{\mathrm{los}}) [\bar{\phi}'(x-\frac{\Delta x}{2})e^{-\tau_{\mathrm{dust}}} - \bar{\phi}'(x+\frac{\Delta x}{2})], \nonumber \\ 
        &=  -e^{-\tau_{\mathrm{dust}}} S_{\nu_0} \tau_{\nu_0} x_Z (\hat{\boldsymbol{\phi}}\cdot \hat{\boldsymbol{n}}_{\mathrm{los}}) [\bar{\phi}'(x-\frac{\Delta x}{2}) - \bar{\phi}'(x+\frac{\Delta x}{2})] \nonumber \\ &+ (1-e^{-\tau_{\mathrm{dust}}})  S_{\nu_0} \tau_{\nu_0} x_Z (\hat{\boldsymbol{\phi}}\cdot \hat{\boldsymbol{n}}_{\mathrm{los}}) \bar{\phi}'(x+\frac{\Delta x}{2}), \\
        &\simeq  -e^{-\tau_{\mathrm{dust}}} S_{\nu_0} \tau_{\nu_0} \Delta x x_Z (\hat{\boldsymbol{\phi}}\cdot \hat{\boldsymbol{n}}_{\mathrm{los}}) \bar{\phi}''(x) \nonumber \\ &+ (1-e^{-\tau_{\mathrm{dust}}})  S_{\nu_0} \tau_{\nu_0} x_Z (\hat{\boldsymbol{\phi}}\cdot \hat{\boldsymbol{n}}_{\mathrm{los}}) \bar{\phi}'(x).
\end{align}
It can then be recognized that due to the dusty midplane layer, the total intensity is adjusted to 
\begin{subequations}
\begin{align}    
I_{\nu}^{\mathrm{c.s.}} \to \frac{1+e^{-\tau_{\mathrm{dust}}}}{2} I_{\nu}, 
\end{align}
while the circular polarization due to the regular Zeeman effect and the velocity shift, are adjusted to
\begin{align}    
V_{\nu}^{\mathrm{reg},\ \mathrm{dust}} &\to \frac{1-e^{-\tau_{\mathrm{dust}}}}{2} V_{\nu}^{\mathrm{reg}}, \\
V_{\nu}^{\mathrm{vel},\ \mathrm{dust}} &\to e^{-\tau_{\mathrm{dust}}} V_{\nu}^{\mathrm{vel}}.
\end{align}
Adjusting the estimates for the emergent polarization fraction using these factors yields Eq.~(\ref{eq:pV_dust}).

\end{subequations}

\section{Zeeman signatures of CN from a TW Hya like disk in ALMA band $3$, $6$ and $7$}
\label{sec:ap_C}
\begin{table*}[]
    \caption{Estimates of the Zeeman induced circular polarization fractions, linear polarization fractions and Zeeman broadening of CN transitions excited in a TW Hya like disk that are relevant to ALMA polarization measurements. Estimates are given for deprojected distance $r_c=50$ AU and azimuthal angles $\phi'=0^o$ and $\phi'=90^0$. For circular polarization estimates, a dusty midplane layer is assumed ($\tau_{\mathrm{dust}}=0.7,\ 1.0,\ 2.0$ for band 3, 6 and 7, respectively), while for the linear polarization and Zeeman broadening, an optically thin midplane is assumed.}
\label{tab:CN_obs_all}
    \centering
    \begin{tabular}{l l l l l l c c c c c c c}
    \hline \hline
        &     &     &      &      &      &                        &             & $\phi'=0^o$                 &             &              & $\phi'=90^o$                  &             \\
    $N$ & $J$ & $F$ & $N'$ & $J'$ & $F'$ & $\nu \ \mathrm{(GHz)}$ & $p_V\ (\%)$ & $\Delta v_Z \ (\mathrm{m/s})$ & $p_l\ (\%)$ &  $p_V\ (\%)$ & $\Delta v_Z \ (\mathrm{m/s})$ & $p_l\ (\%)$ \\
    \hline

$1$ & $0.5$ & $0.5$ & $0$ & $0.5$ & $1.5$ & $113.14416$ & $2.24$ & $27.21$ & $7.30$ & $1.06$ & $26.96$ & $7.38$ \\
$1$ & $0.5$ & $0.5$ & $0$ & $0.5$ & $0.5$ & $113.12337$ & $-0.64$ & $9.58$ & $1.83$ & $-0.30$ & $9.64$ & $1.85$ \\
$1$ & $0.5$ & $1.5$ & $0$ & $0.5$ & $1.5$ & $113.19128$ & $0.64$ & $82.63$ & $13.43$ & $0.30$ & $83.10$ & $13.58$ \\
$1$ & $0.5$ & $1.5$ & $0$ & $0.5$ & $0.5$ & $113.17049$ & $-0.31$ & $3.86$ & $0.01$ & $-0.15$ & $3.86$ & $0.01$ \\
$1$ & $1.5$ & $2.5$ & $0$ & $0.5$ & $1.5$ & $113.49097$ & $0.57$ & $6.62$ & $0.26$ & $0.27$ & $6.61$ & $0.26$ \\
$1$ & $1.5$ & $1.5$ & $0$ & $0.5$ & $1.5$ & $113.50891$ & $1.65$ & $16.34$ & $3.50$ & $0.79$ & $16.22$ & $3.54$ \\
$1$ & $1.5$ & $1.5$ & $0$ & $0.5$ & $0.5$ & $113.48812$ & $2.22$ & $46.10$ & $6.34$ & $1.06$ & $45.88$ & $6.41$ \\
$1$ & $1.5$ & $0.5$ & $0$ & $0.5$ & $1.5$ & $113.52043$ & $1.59$ & $15.43$ & $3.63$ & $0.76$ & $15.31$ & $3.67$ \\
$1$ & $1.5$ & $0.5$ & $0$ & $0.5$ & $0.5$ & $113.49964$ & $0.64$ & $32.17$ & $9.08$ & $0.30$ & $32.49$ & $9.18$ \\
$2$ & $1.5$ & $0.5$ & $1$ & $0.5$ & $0.5$ & $226.66369$ & $-0.38$ & $2.38$ & $0.46$ & $-0.15$ & $2.40$ & $0.46$ \\
$2$ & $1.5$ & $0.5$ & $1$ & $0.5$ & $1.5$ & $226.61657$ & $-0.19$ & $0.96$ & $0.00$ & $-0.07$ & $0.96$ & $0.00$ \\
$2$ & $1.5$ & $0.5$ & $1$ & $1.5$ & $1.5$ & $226.29894$ & $1.33$ & $11.59$ & $1.60$ & $0.53$ & $11.53$ & $1.61$ \\
$2$ & $1.5$ & $0.5$ & $1$ & $1.5$ & $0.5$ & $226.28742$ & $0.38$ & $8.09$ & $2.28$ & $0.15$ & $8.17$ & $2.31$ \\
$2$ & $1.5$ & $1.5$ & $1$ & $0.5$ & $0.5$ & $226.67931$ & $-0.72$ & $2.88$ & $0.50$ & $-0.29$ & $2.86$ & $0.50$ \\
$2$ & $1.5$ & $1.5$ & $1$ & $0.5$ & $1.5$ & $226.63219$ & $-0.44$ & $0.78$ & $0.18$ & $-0.18$ & $0.78$ & $0.18$ \\
$2$ & $1.5$ & $1.5$ & $1$ & $1.5$ & $2.5$ & $226.33250$ & $1.58$ & $16.92$ & $2.22$ & $0.63$ & $16.84$ & $2.24$ \\
$2$ & $1.5$ & $1.5$ & $1$ & $1.5$ & $1.5$ & $226.31454$ & $0.17$ & $15.69$ & $2.69$ & $0.07$ & $15.78$ & $2.72$ \\
$2$ & $1.5$ & $1.5$ & $1$ & $1.5$ & $0.5$ & $226.30304$ & $-1.11$ & $12.47$ & $0.90$ & $-0.44$ & $12.44$ & $0.91$ \\
$2$ & $1.5$ & $2.5$ & $1$ & $0.5$ & $1.5$ & $226.65956$ & $-0.44$ & $0.64$ & $0.20$ & $-0.17$ & $0.63$ & $0.20$ \\
$2$ & $1.5$ & $2.5$ & $1$ & $1.5$ & $2.5$ & $226.35987$ & $0.14$ & $26.39$ & $4.33$ & $0.05$ & $26.54$ & $4.37$ \\
$2$ & $1.5$ & $2.5$ & $1$ & $1.5$ & $1.5$ & $226.34193$ & $-1.35$ & $15.49$ & $1.47$ & $-0.54$ & $15.44$ & $1.48$ \\
$2$ & $2.5$ & $3.5$ & $1$ & $1.5$ & $2.5$ & $226.87478$ & $0.25$ & $0.70$ & $0.04$ & $0.10$ & $0.70$ & $0.04$ \\
$2$ & $2.5$ & $2.5$ & $1$ & $0.5$ & $1.5$ & $227.19182$ & $1.34$ & $11.86$ & $1.62$ & $0.53$ & $11.80$ & $1.63$ \\
$2$ & $2.5$ & $2.5$ & $1$ & $1.5$ & $2.5$ & $226.89213$ & $0.65$ & $1.53$ & $0.41$ & $0.26$ & $1.52$ & $0.42$ \\
$2$ & $2.5$ & $2.5$ & $1$ & $1.5$ & $1.5$ & $226.87419$ & $0.44$ & $0.95$ & $0.18$ & $0.17$ & $0.94$ & $0.19$ \\
$2$ & $2.5$ & $1.5$ & $1$ & $1.5$ & $2.5$ & $226.90536$ & $0.48$ & $1.22$ & $0.22$ & $0.19$ & $1.22$ & $0.22$ \\
$2$ & $2.5$ & $1.5$ & $1$ & $1.5$ & $1.5$ & $226.88742$ & $0.90$ & $2.82$ & $0.82$ & $0.36$ & $2.79$ & $0.82$ \\
$2$ & $2.5$ & $1.5$ & $1$ & $1.5$ & $0.5$ & $226.87590$ & $0.72$ & $3.03$ & $0.49$ & $0.29$ & $3.02$ & $0.49$ \\
$3$ & $2.5$ & $1.5$ & $2$ & $1.5$ & $0.5$ & $340.03541$ & $-0.38$ & $0.62$ & $0.15$ & $-0.15$ & $0.61$ & $0.15$ \\
$3$ & $2.5$ & $1.5$ & $2$ & $1.5$ & $1.5$ & $340.01963$ & $-0.55$ & $0.65$ & $0.14$ & $-0.16$ & $0.65$ & $0.14$ \\ 
$3$ & $2.5$ & $1.5$ & $2$ & $1.5$ & $2.5$ & $339.99226$ & $-0.19$ & $0.27$ & $0.01$ & $-0.05$ & $0.27$ & $0.01$ \\ 
$3$ & $2.5$ & $1.5$ & $2$ & $2.5$ & $2.5$ & $339.46000$ & $1.47$ & $8.21$ & $0.95$ & $0.42$ & $8.18$ & $0.96$ \\ 
$3$ & $2.5$ & $1.5$ & $2$ & $2.5$ & $1.5$ & $339.44678$ & $0.13$ & $10.53$ & $1.82$ & $0.04$ & $10.59$ & $1.84$ \\ 
$3$ & $2.5$ & $2.5$ & $2$ & $1.5$ & $1.5$ & $340.03541$ & $-0.35$ & $0.23$ & $0.07$ & $-0.10$ & $0.23$ & $0.07$ \\ 
$3$ & $2.5$ & $2.5$ & $2$ & $1.5$ & $2.5$ & $340.00813$ & $-0.39$ & $0.27$ & $0.08$ & $-0.11$ & $0.27$ & $0.08$ \\ 
$3$ & $2.5$ & $2.5$ & $2$ & $2.5$ & $3.5$ & $339.49321$ & $1.53$ & $8.97$ & $1.04$ & $0.44$ & $8.93$ & $1.05$ \\ 
$3$ & $2.5$ & $2.5$ & $2$ & $2.5$ & $2.5$ & $339.47590$ & $0.08$ & $10.54$ & $1.73$ & $0.02$ & $10.60$ & $1.75$ \\ 
$3$ & $2.5$ & $2.5$ & $2$ & $2.5$ & $1.5$ & $339.46264$ & $-1.38$ & $8.50$ & $0.78$ & $-0.39$ & $8.48$ & $0.79$ \\ 
$3$ & $2.5$ & $3.5$ & $2$ & $1.5$ & $2.5$ & $340.03155$ & $-0.25$ & $0.13$ & $0.03$ & $-0.07$ & $0.13$ & $0.03$ \\ 
$3$ & $2.5$ & $3.5$ & $2$ & $2.5$ & $3.5$ & $339.51664$ & $0.07$ & $12.65$ & $2.05$ & $0.02$ & $12.72$ & $2.07$ \\ 
$3$ & $2.5$ & $3.5$ & $2$ & $2.5$ & $2.5$ & $339.49929$ & $-1.44$ & $8.85$ & $0.87$ & $-0.41$ & $8.82$ & $0.88$ \\
$3$ & $3.5$ & $4.5$ & $2$ & $2.5$ & $3.5$ & $340.24777$ & $0.18$ & $0.17$ & $0.01$ & $0.05$ & $0.17$ & $0.01$ \\
$3$ & $3.5$ & $3.5$ & $2$ & $2.5$ & $3.5$ & $340.26495$ & $0.44$ & $0.35$ & $0.10$ & $0.13$ & $0.35$ & $0.10$ \\
$3$ & $3.5$ & $3.5$ & $2$ & $2.5$ & $2.5$ & $340.24777$ & $0.25$ & $0.24$ & $0.03$ & $0.07$ & $0.24$ & $0.03$ \\
$3$ & $3.5$ & $2.5$ & $2$ & $2.5$ & $3.5$ & $340.27912$ & $0.29$ & $0.26$ & $0.04$ & $0.08$ & $0.26$ & $0.04$ \\
$3$ & $3.5$ & $2.5$ & $2$ & $2.5$ & $2.5$ & $340.26177$ & $0.57$ & $0.57$ & $0.18$ & $0.16$ & $0.57$ & $0.18$ \\
$3$ & $3.5$ & $2.5$ & $2$ & $2.5$ & $1.5$ & $340.24854$ & $0.35$ & $0.51$ & $0.05$ & $0.10$ & $0.51$ & $0.06$ \\
\hline
    \end{tabular}
\end{table*}

\end{appendix}

\end{document}